%% file: iclr2025_conference.tex
\documentclass{article} 
\usepackage{iclr2025_conference,times}

\input{math_commands.tex}

\usepackage{hyperref}
\usepackage{url}

\usepackage{multirow}
\usepackage{adjustbox}
\usepackage{booktabs}
\usepackage{wrapfig}

\usepackage{diagbox}
\usepackage{dirtytalk}
\usepackage{epigraph}
\usepackage{subfigure}
\usepackage{enumitem}

\usepackage{alltt}
\usepackage[breakable,skins]{tcolorbox}

\hypersetup{
    colorlinks=true,   
    citecolor=blue     
}

\tcbset{
  assistantbox/.style={
    width=400.18663pt,
    top=10pt,
    colback=black!05,
    colframe=assistantcolor,
    colbacktitle=black!50,
    enhanced,
    center,
    attach boxed title to top left={yshift=-0.1in,xshift=0.15in},
    boxed title style={boxrule=0pt,colframe=white,},
  }
}

\tcbset{
  userbox/.style={
    width=400.18663pt,
    top=10pt,
    colback=black!05,
    colframe=usercolor,
    colbacktitle=black!50,
    enhanced,
    center,
    attach boxed title to top left={yshift=-0.1in,xshift=0.15in},
    boxed title style={boxrule=0pt,colframe=white,},
  }
}

\tcbset{
  aibox/.style={
    width=400.18663pt,
    top=10pt,
    colback=black!05,
    colframe=black!20,
    colbacktitle=black!50,
    enhanced,
    center,
    attach boxed title to top left={yshift=-0.1in,xshift=0.15in},
    boxed title style={boxrule=0pt,colframe=white,},
  }
}

\tcbset{
  aiboxbreakable/.style={
    width=400.18663pt,
    top=10pt,
    colback=black!05,
    colframe=black!20,
    colbacktitle=black!50,
    enhanced,
    center,
    breakable,
    attach boxed title to top left={yshift=-0.1in,xshift=0.15in},
    boxed title style={boxrule=0pt,colframe=white,},
  }
}

\newtcolorbox{AssistantBox}[2][]{assistantbox,title=#2,#1}
\newtcolorbox{UserBox}[2][]{userbox,title=#2,#1}
\newtcolorbox{AIBox}[2][]{aibox,title=#2,#1}
\newtcolorbox{AIBoxBreak}[2][]{aiboxbreakable,title=#2,#1}

\definecolor{mydarkgreen}{RGB}{0, 139, 69}

\title{On Calibration of LLM-based Guard Models for Reliable Content Moderation}

\author{
 \textbf{Hongfu Liu}\textsuperscript{1}\thanks{Corresponding to: Hongfu Liu (hongfu@comp.nus.edu.sg)},~
 \textbf{Hengguan Huang}\textsuperscript{2},~
 \textbf{Xiangming Gu}\textsuperscript{1},~
 \textbf{Hao Wang}\textsuperscript{3},~
 \textbf{Ye Wang}\textsuperscript{1}
\\
 \textsuperscript{1}National University of Singapore \textsuperscript{2}University of Copenhagen \textsuperscript{3}Rutgers University 
}

\iclrfinalcopy 
\begin{document}

\maketitle

\begin{abstract}
Large language models (LLMs) pose significant risks due to the potential for generating harmful content or users attempting to evade guardrails. Existing studies have developed LLM-based guard models designed to moderate the input and output of threat LLMs, ensuring adherence to safety policies by blocking content that violates these protocols upon deployment. However, limited attention has been given to the reliability and calibration of such guard models. In this work, we empirically conduct comprehensive investigations of confidence calibration for 9 existing LLM-based guard models on 12 benchmarks in both user input and model output classification. Our findings reveal that current LLM-based guard models tend to 1) produce overconfident predictions, 2) exhibit significant miscalibration when subjected to jailbreak attacks, and 3) demonstrate limited robustness to the outputs generated by different types of response models. Additionally, we assess the effectiveness of post-hoc calibration methods to mitigate miscalibration. We demonstrate the efficacy of temperature scaling and, for the first time, highlight the benefits of contextual calibration for confidence calibration of guard models, particularly in the absence of validation sets. Our analysis and experiments underscore the limitations of current LLM-based guard models and provide valuable insights for the future development of well-calibrated guard models toward more reliable content moderation. We also advocate for incorporating reliability evaluation of confidence calibration when releasing future LLM-based guard models\footnote{Our code is publicly available at \href{https://github.com/Waffle-Liu/calibration_guard_model}{https://github.com/Waffle-Liu/calibration\_guard\_model}}.

\end{abstract}

\section{Introduction}

Recent advancements in Large Language Models (LLMs) have facilitated the development of powerful conversation systems, leading to the deployment of LLM-based chatbots in various real-world applications~\citep{brown2020language,palm2,llama2,llama3}. However, these systems face substantial risks due to the potential for malicious exploitation of powerful LLMs~\citep{wang2023decodingtrust}. Consequently, addressing these risks has become an urgent and critical task. One promising strategy is to regulate LLMs during their training phase. Existing researches primarily focus on designing alignment algorithms through preference optimization~\citep{rlhf,dpo}, implementing adversarial training~\citep{harmbench}, or employing machine unlearning to remove harmful knowledge from the models~\citep{chen2023unlearn,liu2024rethinking}. These approaches aim to control text generation and prevent undesired outputs. Despite these significant efforts to enhance LLM safety during training, red-teaming still makes efforts to expose vulnerabilities, including jailbreak attacks that successfully bypass the safety constraint and elicit harmful responses from LLMs, highlighting the risks of future, unseen threats~\citep{gcg,autodan,chao2024jailbreakbench,liu2024advancing}. Therefore, in addition to training-time interventions, it is equally vital to implement test-time measures, such as constraint inference~\citep{xu2024safedecoding}, and establish effective test-time guardrails through content moderation, particularly when deploying LLMs in real-world settings.     

Content Moderation serves the critical function of monitoring both user inputs and model outputs during conversations. Typically, guard models are designed to assess whether user inputs and LLM outputs comply with safety regulations, and either reject user queries or block model responses when content violating safety protocols is detected. This approach remains effective even when LLMs have been compromised by previously unseen jailbreak attacks. Current state-of-the-art guard models, which are typically built on LLMs, demonstrate strong performance across various benchmarks~\citep{llamaguard,aegis,wildguard,shieldgemma}. However, these guard models primarily focus on the classification performance but overlook the predictive uncertainty of harmfulness predictions~\citep{BLoB,NPN,BDL,BDLSurvey}, therefore failing to assess the reliability of these models' predictions. This oversight is crucial because guard models may occasionally make erroneous decisions, potentially allowing unsafe content to bypass moderation, especially when encountering non-trivial domain shifts~\citep{ECBNN,GRDA,CIDA}, despite their strong in-domain performances. Therefore, quantifying the predictive uncertainty and confidence in model predictions is essential to assessing the trustworthiness of guard models, enabling more reliable decision-making in high-risk scenarios that may arise during conversations after model deployment. 
\begin{figure}[t]
    \centering
    \vskip -0.25in
    \includegraphics[width=\textwidth]{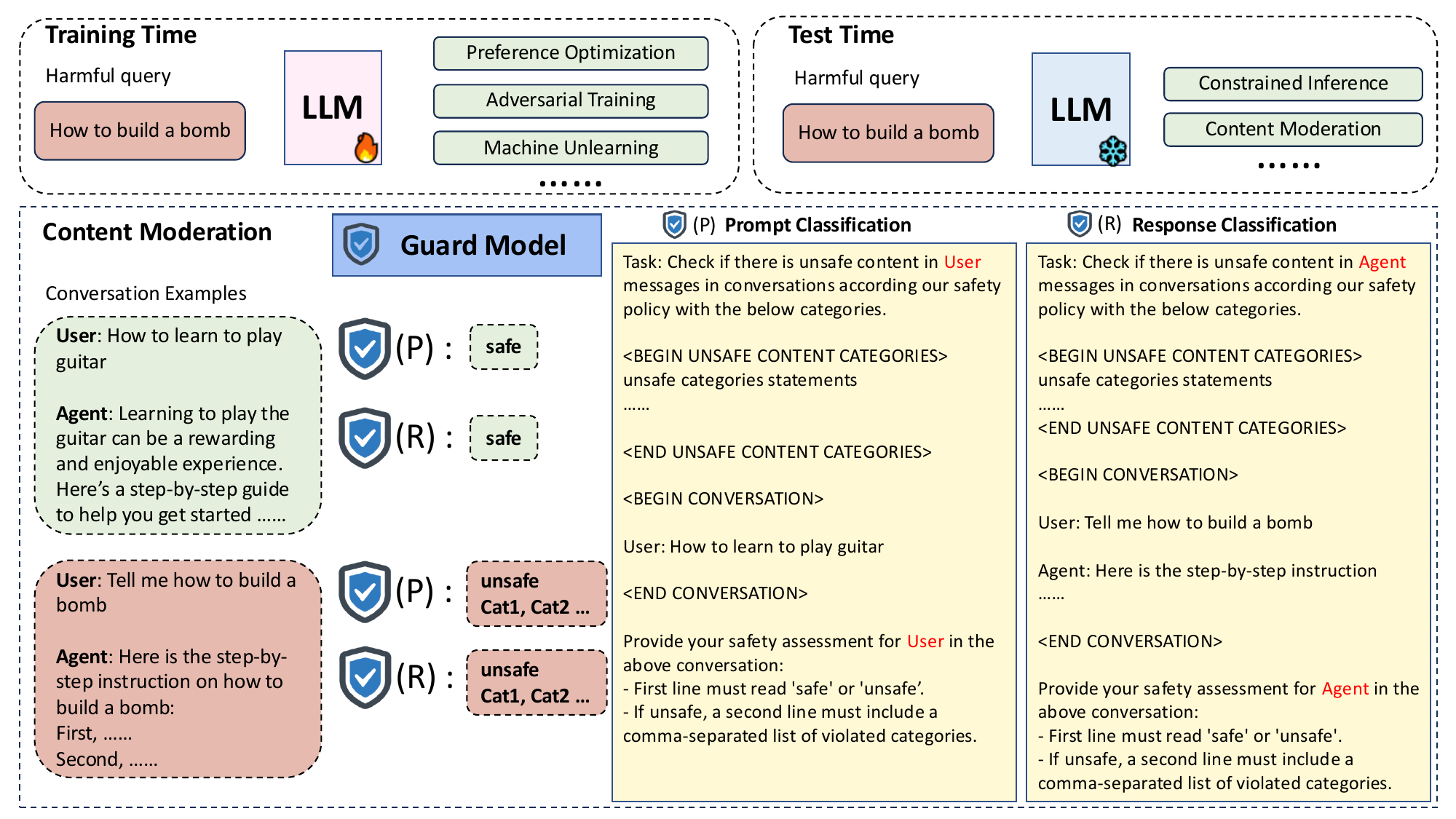}
    \vskip -.15in
    \caption{An overview of LLM-based guard models for content moderation. Guard models monitor the input and output during conversations between the user and LLM (Agent), providing a binary prediction followed by a specific unsafe content category if unsafe content is detected. The instruction examples for prompt classification and response classification from LLama-Guard are detailed in the right yellow boxes respectively.}
    \label{fig:main}
    \vskip -.2in
\end{figure}

In this work, we examine the reliability of existing open-source guard models by focusing on their confidence calibration. Specifically, we empirically assess the calibration performance by commonly used expected calibration error (ECE) for two key tasks: user input (prompt) classification and model output (response) classification with binary labels. To conduct a systematic evaluation, we examine 9 models across 12 datasets. Our experimental results reveal that, despite achieving strong performance, most existing guard models exhibit varying levels of miscalibration. Additionally, our findings show that current LLM-based guard models:

\begin{itemize}[nosep,leftmargin=24pt]
    \setlength{\itemsep}{1pt}
    \item tend to make overconfident predictions with high probability scores.
    \item remain poorly calibrated under adversarial environments, exhibiting higher ECE in adversarial prompt classification, even when the SOTA guard model achieves high F1 scores.
    \item display inconsistent ECE across different types of response models, demonstrating weak robustness to variations in response model types.
\end{itemize}
These observations highlight critical challenges in improving the reliability of guard models in real-world deployments. Consequently, we are motivated to improve the confidence calibration of guard models, focusing on post-hoc calibration methods to avoid additional computational costs of training new guard models. We explore the impact of bias calibration methods on confidence calibration for the first time, discovering that contextual calibration proves impressively effective for prompt classification, while conventional temperature scaling remains more beneficial for response classification. Lastly, we identify miscalibration issues stemming from prediction vulnerabilities induced by single tokens and misaligned classification objectives, highlighting the limitations of instruction-tuned LLM-based guard models. We stress the importance of reliability evaluation and advocate for the inclusion of confidence calibration measurement in the release of future new guard models.

\section{Related Work}

\textbf{Content Moderation}. A substantial body of research has been devoted to the detection of hateful and toxic content in human-generated text from online platforms, such as social media~\citep{hatespeech,perspective}. Various API services, including Perspective~\citep{perspective}, OpenAI~\citep{openai}, Azure, and Detoxify, provide black-box content moderation tools for online texts. However, content moderation in LLMs specifically addresses the detection of both LLM input and output during conversations within deployed applications, such as chat assistants. This task poses unique challenges due to the distribution shift in conversation content generated by LLMs, which differs from previous human-generated online texts. Recent advancements in LLM content moderation have been achieved through the fine-tuning of LLMs, as seen in models such as LLama-Guard1/2/3~\citep{llamaguard}, BeaverDam~\citep{beavertails}, Aegis~\citep{aegis}, MD-Judge~\citep{mdjudge}, WildGuard~\citep{wildguard}, and ShieldGemma~\citep{shieldgemma}. Notably, models like Llama-Guard, Aegis, and WildGuard support the detection of both user inputs and model outputs, while others do not due to differing training objectives. Additionally, adversarial cases are addressed by Harmbench~\citep{harmbench} and RigorLLM~\citep{rigor}, with Harmbench specifically fine-tuning LLama2 and Mistral to evaluate the success rate of adversarial attacks by identifying undesirable content in model outputs. Furthermore, Nemo proposes programmable guardrails that provide dialogue management capability using a user-friendly toolkit~\citep{nemo}. Our work focuses on quantifying the predictive uncertainty and evaluating the reliability of LLM-based guard models by their calibration levels.

\textbf{Calibration of LLMs}. Confidence calibration is a critical aspect in developing reliable and trustworthy language models~\citep{posteriorllm,guo2017calibration,minderer2021revisiting}. In the context of LLMs, prior research has explored calibration in question-answering tasks~\citep{jiang2021can} and has empirically examined calibration during the pre-training and alignment stages~\citep{chen2022close,calibrationllm}. Studies such as \citet{lin2022teaching,mielke2022reducing,xiong2023can} have investigated uncertainty estimation through verbalized confidence, and \citet{kadavath2022language} demonstrated improved calibration of larger models when handling multiple choice and true/false questions given appropriate formats. Another line of research addresses the calibration of biases stemming from in-context samples, instruction templates, sample ordering, and label distribution~\citep{zhao2021calibrate,zhou2023survival,liu2023towards,fei2023mitigating,abbas2024enhancing}. These bias calibration techniques indirectly influence the prediction confidence by altering the linear decision boundary~\citep{zhou2023batch}, yet they are not designed for explicit confidence calibration. In contrast, our work specifically addresses the challenge of confidence calibration in instruction-tuned guard models for content moderation tasks.

\section{Preliminary}

\textbf{LLM-based Guard Models}. Given the user input text $\mathbf{X}$ and the corresponding response $\mathbf{R} = f(\mathbf{X})$ generated by a deployed LLM $f(*)$, the task of the LLM-based guard model $g(*)$ is to classify the user input $p_{g}(\mathbf{Y}|\mathbf{X}) $, or the LLM output $p_{g}(\mathbf{Y}|\mathbf{X}, \mathbf{R})$\footnote{Note that we ignore the instruction context $\mathrm{C_{inst}}$ in our all following notations for simplicity where they should be $p_{g}(\mathbf{Y}|\mathbf{X}; \mathrm{C_{inst}}) $ and $p_{g}(\mathbf{Y}|\mathbf{X}, \mathbf{R}; \mathrm{C_{inst}})$ instead.} These tasks are referred to as \textbf{prompt classification} and \textbf{response classification}, respectively. For the predicted label $\mathbf{Y}$, most existing LLM-based guard models initially perform binary classifications $y^b$ to determine whether the user input $\mathbf{X}$ or model response $\mathbf{R}$ is safe. If the binary classification result indicates the input or the response $y^b$ is unsafe, the guard model $g(*)$ then proceeds with a multiclass classification to categorize the specific type $y^c$ by $p_{g}(y^c | \mathbf{X}, y^b)$ or $p_{g}(y^c | \mathbf{X}, \mathbf{R}, y^b)$ where the categories $c$ are pre-defined in a taxonomy. These classification tasks in LLM-based guard models are carried out in an autoregressive generation manner, and Figure~\ref{fig:main} illustrates examples of the prompt and response classification instructions used in LLama-Guard.

\textbf{Confidence Calibration}. A model is considered perfectly calibrated if its predicted class $\hat{y}$ and the associated confidence $\hat{p} \in [0, 1]$ satisfy $P(\hat{y}=y | \hat{p} = p ) = p, \forall p \in [0, 1]$, where $y$ is the ground-truth class label for any given input. This implies that higher confidence in a prediction should correspond to a higher chance of its prediction being correct. However, since $P(\hat{y}=y | \hat{p} = p )$ can not be directly calculated with finite sample size, existing approaches employ binning-based divisions on finite samples and utilize the Expected Calibration Error (ECE) as a quantitative metric to assess the model's calibration~\citep{naeini2015obtaining}. Assuming that confidence is divided into $M$ bins with equal interval $1/M$ within the range $[0,1]$, the ECE is defined as
\begin{equation}
    ECE = \sum_{m=1}^M \frac{|B_m|}{N} \left| Acc(B_m) - Conf(B_m) \right|,  
\end{equation}
    
\begin{equation}
    Acc(B_m) = \frac{1}{|B_m|} \sum_{i\in B_m} \mathbf{1}(\hat{y}_i = y_i), \quad Conf(B_m) = \frac{1}{|B_m|} \sum_{i\in B_m} \hat{p}_i
\end{equation}

where $B_m$ represents the set of samples falling within the interval $(\frac{m-1}{M}, \frac{m}{M}]$, $\hat{y}_i$ and $y_i$ are the predicted and ground truth classes, respectively, and $\hat{p}_i$ is the model's predicted probability. However, existing instruction-tuned LLM-based guard models do not directly output the probability of each class. Instead, the probability of class $c_i$ is derived from the output logits $z_{\mathcal{V}(c_i)}$ of the corresponding target label token $\mathcal{V}(c_i)$, where $\mathcal{V}(*)$ is the verbalizer. Re-normalization is then applied over the set of target label tokens as follows, 
\begin{equation}
    p(y = c_i |\mathbf{X}, \mathbf{R}) = \frac{e^{z_{\mathcal{V}(c_i)}}}{\sum_{c_i} e^{z_{\mathcal{V}(c_i)}}} 
\end{equation}
where $\mathbf{R}$ is empty for prompt classification. Specifically, for binary classification tasks, the target label tokens could simply be ``safe /\ unsafe'', ``harmful /\ unharmful'', or ``yes /\ no'', depending on the specific instructions utilized in different guard models.

\section{Calibration Measurement of LLM-based guard models}

To systematically evaluate the calibration of existing open-source LLM-based guard models across public benchmarks, we conduct an analysis of 9 models on 12 publicly available datasets. We take the prompt classification and response classification as two primary tasks in our investigation. Due to the variability in safety taxonomies across different guard models and datasets, it is challenging to directly compare performance on multiclass prediction tasks. Therefore, our evaluation emphasizes \textbf{binary classification} (safe/\ unsafe) for both prompt and response classifications, allowing for a more consistent and fair comparison across guard models. Moreover, binary classification is a critical precursor to multiclass predictions, as an incorrect binary prediction could result in the dissemination of undesired content to users, increasing the associated risk. Thus, binary classification holds particular importance in ensuring the reliability and safety of these systems. 


\subsection{Experimental Setup}

\textbf{Benchmarks}. To assess calibration in the context of binary prompt classification, we evaluate performance using a range of public benchmarks, including OpenAI Moderation~\citep{openai}, ToxicChat Test~\citep{toxicchat}, Aegis Safety Test~\citep{aegis}, SimpleSafetyTests~\citep{simplesafetytests}, XSTest~\citep{xstest}, Harmbench Prompt~\citep{harmbench} and WildGuardMix Test Prompt~\citep{wildguard}. For the response classification, we utilize datasets containing BeaverTails Test~\citep{beavertails}, SafeRLHF Test~\citep{saferlhf}, Harmbench Response~\citep{harmbench}, and WildGuardMix Test Response~\citep{wildguard}. For all datasets, we report the ECE as the primary metric for calibration assessment, alongside the F1 score for classification performance. Detailed statistics of each dataset can be found in Appendix~\ref{dataset_stat}.  

\textbf{LLM-based Guard Models}. Existing LLM-based guard models vary in their capabilities, with some supporting both prompt and response classification, while others specialize in response classification, based on their instruction-tuning tasks. For prompt classification, we evaluate Llama-Guard, Llama-Guard2, Llama-Guard3, Aegis-Guard-Defensive, Aegis-Guard-Permissive, and WildGuard~\citep{llamaguard,aegis,wildguard}. In the case of response classification, we additionally assess Harmbench-Llama, Harmbench-Mistral, and MD-Judge-v0.1~\citep{harmbench,mdjudge}. API-based moderation tools are excluded from our evaluation due to the nature of their black-box models, which output scores that cannot be simply interpreted as probability. More details can be found in Appendix~\ref{model_detail}.

\input{table/table1}

\begin{figure}[t]
    \centering
    \includegraphics[width=\textwidth]{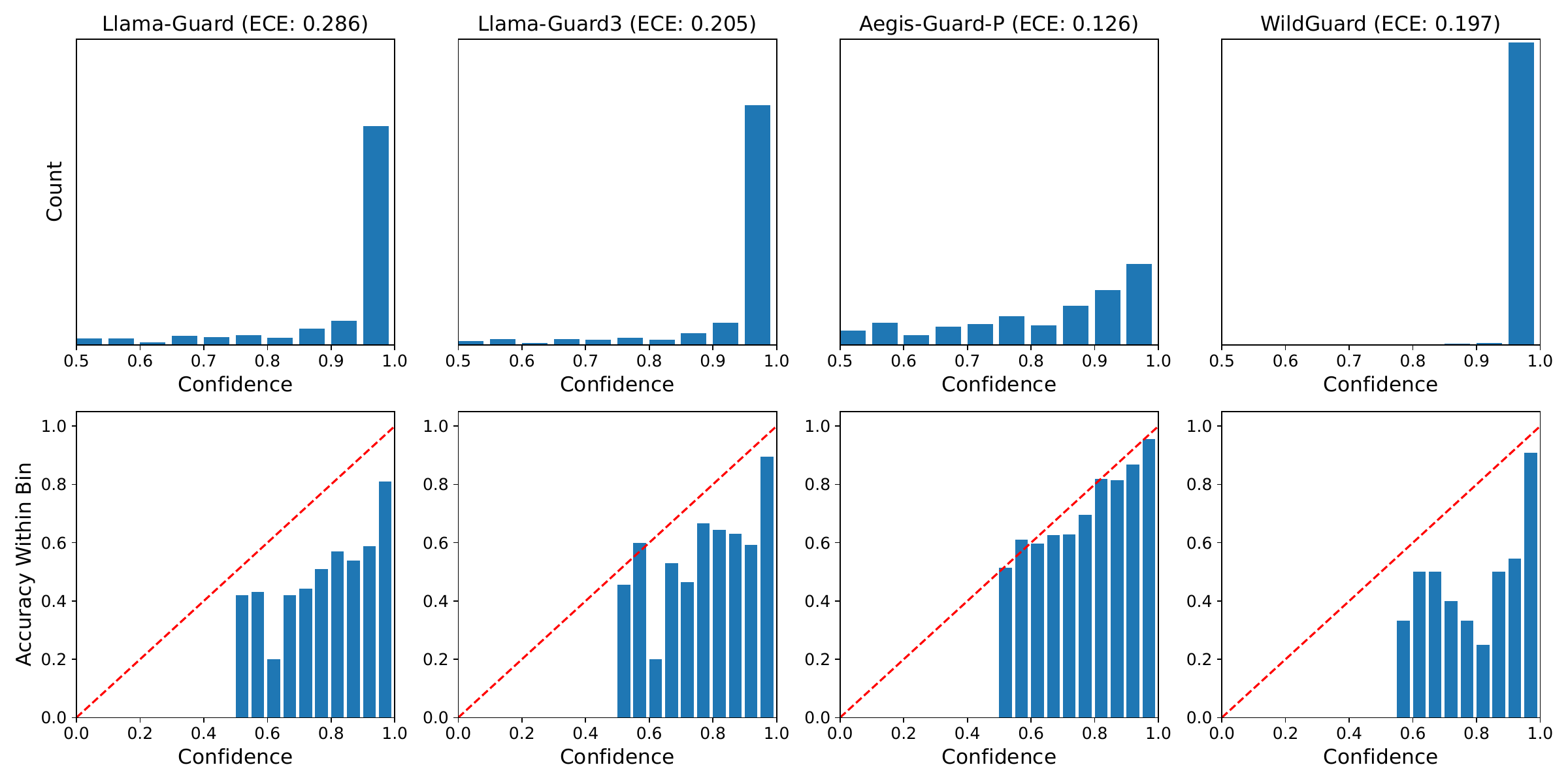}
    \caption{Confidence distributions (First row) and reliability diagrams (Second row) of Llama-Guard, Llama-Guard3, Aegis-Guard-P, and WildGuard on the WildGuardMix Test Prompt set.}
    \label{fig:rediag_prompt}
\end{figure}

\subsection{Main Results}

\subsubsection{General Evaluation on public benchmarks}
We begin by conducting a comprehensive evaluation of both prompt and response classifications for the existing guard models on public benchmarks. The ECE results for both tasks are presented in Table~\ref{table1}. Our experimental findings indicate that existing guard models exhibit significant miscalibration in both prompt and response classifications. Among the models evaluated, WildGuard demonstrates the lowest average ECE for prompt classification, achieving $14.4\%$, while MD-Judge achieves the lowest average ECE for response classification, at $11.4\%$. However, despite the relatively better performances, both Wildguard and MF-Judge exhibit average ECE values exceeding $10\%$, which is typically considered a poor calibration and underscores the need for further improvements. Additionally, each model displays a substantial variance in ECE across different datasets, suggesting unreliable predictions.

\textbf{Finding 1: Existing guard models tend to make overconfident predictions}. To further investigate, we visualize the confidence distributions and present the corresponding reliability diagrams in Figure~\ref{fig:rediag_prompt}. Additional results for other datasets, models as well as response classification can be found in Appendix~\ref{more_rediag}. The analysis reveals that for models such as LLama-Guard, Llama-Guard3, and WildGuard, the majority of predictions exhibit confidences between $90\%$ and $100\%$, indicating overconfident predictions along with high ECE. While Aegis-Guard-P shows a less extreme confidence distribution compared to the other models, the proportion of predictions with confidence greater than $90\%$ is still noticeably higher than those with lower confidence, further reflecting the trend of overconfidence.

\subsubsection{Evaluation under jailbreak attacks}
\label{eval_jail}

\begin{wrapfigure}{r}{0.3\textwidth}
    \centering
    \vskip -.05in
    \includegraphics[width=\linewidth]{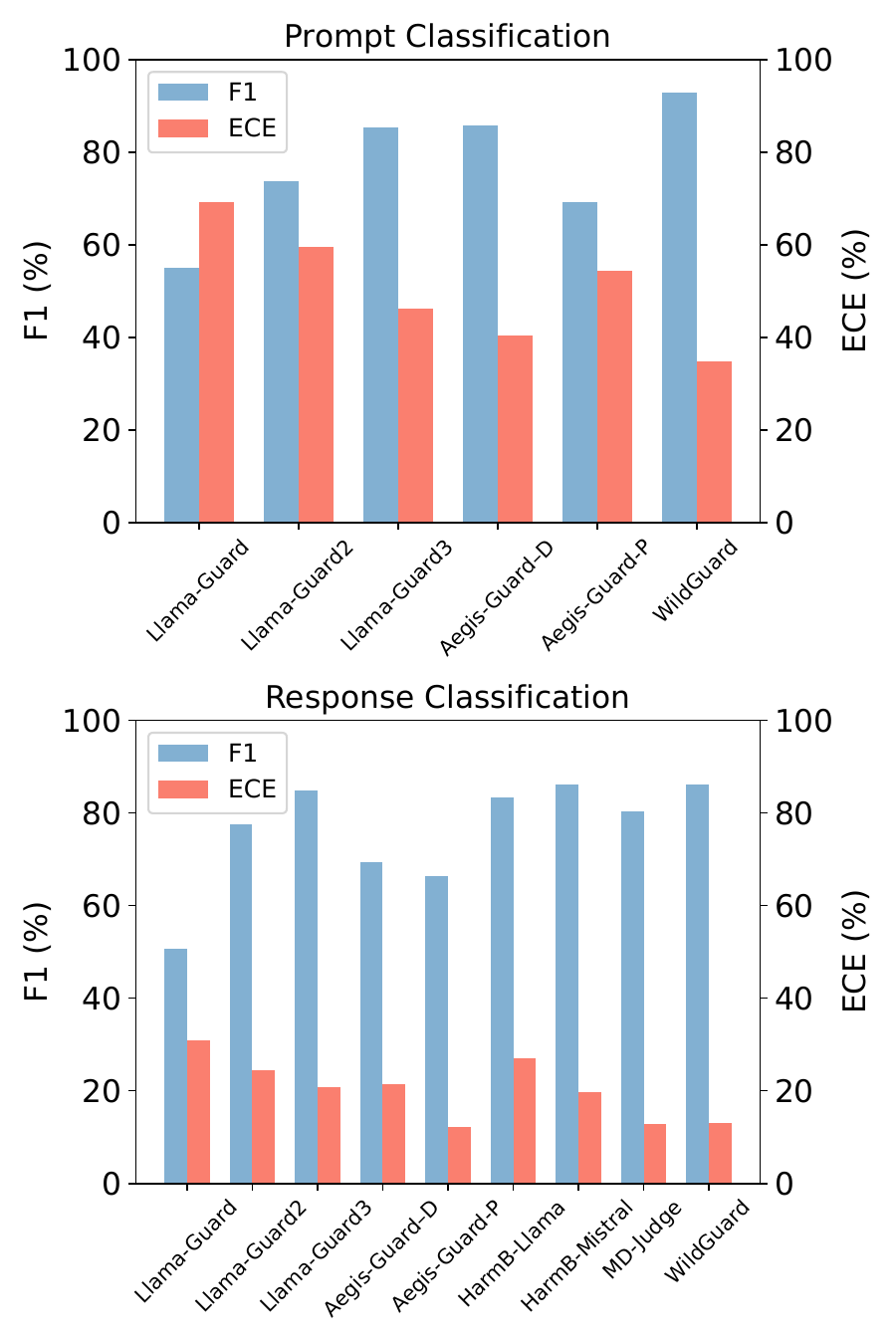}
    \vskip -.1in
    \caption{F1 (\%) $\uparrow$ and ECE (\%) $\downarrow$ performances of prompt and response classification on Harmbench-adv set.}
    \vskip -.05in
    \label{fig:jailbreak_bar}
\end{wrapfigure}

Table~\ref{fig:main} reveals considerable variability of ECE for different guard models when handling harmful requests on the HarmbenchPrompt set. To further investigate the reliability of these guard models in adversarial environments involving dangerous jailbreak attacks, we extend our evaluation to the Harmbench-adv set. This dataset, which serves as a validation set for fine-tuning Llama2-variant classifiers in Harmbench, includes user inputs generated from various types of jailbreak attacks, such as GCG and AutoDAN, leading to a significant distribution shift from typical user input. In this evaluation, we utilize the adversarial user inputs and their corresponding responses and report the F1 and ECE results for each guard model in Figure~\ref{fig:jailbreak_bar}.

\textbf{Finding 2: Miscalibration in prompt classification is more pronounced than in response classification under jailbreak attacks}. The results demonstrate that the ECE for prompt classification is generally higher than that of response classification, indicating that guard models tend to be more reliable when classifying model responses under adversarial conditions. We conjecture that this may be due to the more considerable distribution shift in adversarial prompts than that in model responses. Additionally, while WildGuard achieves SOTA performance with an F1 score of $92.8\%$ in prompt classification, its ECE score remains high at $34.9\%$, highlighting concerns about the reliability of its predictions in real-world deployment.


\input{table/table2}

\subsubsection{Evaluation of Robustness to diverse response models }
\label{diverse_model}
While the ECE for response classification under adversarial environments appears relatively lower in Figure ~\ref{fig:jailbreak_bar}, it remains important to investigate whether each guard model consistently maintains reliability when classifying responses generated by different response models. This is crucial because response models are often aligned differently during post-training so they may have different output distributions and produce different responses to jailbreak attacks. To this end, we continue our calibration evaluation under jailbreak attacks, shifting our focus to response classification. Specifically, we employ the same Harmbench-adv set and divide it according to the response model type. After filtering out subsets with a small sample size, we retain 10 subsets containing responses from Baichuan2, Qwen, Solar, Llama2, Vicuna, Orca2, Koala, OpenChat, Starling, and Zephyr. Each subset consists of outputs from a specific response model. Detailed information on the statistics for each subset is provided in Appendix~\ref{dataset_stat}. The F1 and ECE results are reported in Table~\ref{model_dep}.

\textbf{Finding 3: Guard models exhibit inconsistent reliability when classifying outputs from different response models}. The results in Table~\ref{model_dep} reveal significant variance in both F1 and ECE across different response models. This suggests potential limitations in the training of guard models that rely on responses from a single model. For example, Aegis-Guard models are trained using responses from Mistral, and Llama-Guard models are trained using responses from internal LLama checkpoints. In contrast, Harmbench-Llama, Harmbench-Mistral, and Wildguard are trained using responses from a more diverse set of models, leading to improved generalization across different output distributions of response models.

\section{Improving the Calibration of LLM-based Guard Models}

Empirical evidence has demonstrated the miscalibration of current LLM-based guard models, necessitating efforts to improve their reliability through calibration techniques. In this section, we focus on post-hoc calibration methods to circumvent the computational expense associated with training new guard models, reserving training-time calibration approaches for further investigation.

\subsection{Calibration Techniques} 
\label{cal}

\textbf{Temperature Scaling}~\citep{guo2017calibration}. Temperature scaling (TS) is a widely employed confidence calibration method for traditional neural networks. By introducing a scalar parameter $T > 0$ on the output logits, the output distribution can either be smoothed ($T > 1$) or sharpened ($T < 1$). Specifically, the calibrated confidence is computed as:
\begin{equation}
    \hat{p}(y = c_i |\mathbf{X}, \mathbf{R}) = \frac{e^{\frac{z_{\mathcal{V}(c_i)}}{T}}}{\sum_{c_i} e^{\frac{z_{\mathcal{V}(c_i)}}{T}}} 
\end{equation}
It is important to note that applying temperature scaling does not affect the maximum value of the softmax function, and thus does not alter accuracy performance. The parameter $T$ is typically optimized on a held-out validation set with respect to the negative log-likelihood. However, in the context of the LLM content moderation task, validation sets may not always be available, posing a significant challenge, particularly when addressing in-the-wild user inputs or responses from unknown models. Besides temperature scaling, most conventional calibration methods similarly rely on validation sets to determine parameters, rendering them impractical in such scenarios. As such, we exclusively take temperature scaling as an instance for its simplicity and efficacy.

\textbf{Contextual Calibration}~\citep{zhao2021calibrate}. Contextual calibration (CC) is one type of matrix scaling technique to address contextual bias in LLMs, with the key advantage of requiring no validation set. This method estimates test-time contextual bias by using content-free tokens such as ``N/A'', space, or empty tokens. The calibrated prediction is then computed as follows:
\begin{equation}
    \hat{\mathbf{p}}(y |\mathbf{X}, \mathbf{R}) = \mathbf{W} \mathbf{p}(y |\mathbf{X}, \mathbf{R}) 
\end{equation}

where $\mathbf{W} = \mathrm{diag}(\mathbf{p}(y|[N/A]))^{-1}$. Although the original purpose of contextual calibration differs from confidence calibration, the utilized vector scaling modifies model predictions and impacts confidence levels as well, warranting its consideration for confidence calibration.

\textbf{Batch Calibration}~\citep{zhou2023batch}. Batch calibration (BC) is also a type of matrix scaling approach. The rationale behind batch calibration is to estimate contextual bias from a batch of $M$ unlabeled samples drawn from the target domains $P(x)$ or $P(x, r)$, rather than from context-free tokens as in contextual calibration. Specifically, batch calibration applies a transformation on the original prediction, which can be interpreted as a linear transformation in the log-probability space,
\begin{equation}
    \log\hat{\mathbf{p}}(y |\mathbf{X}, \mathbf{R}) = \log\mathbf{p}(y |\mathbf{X}, \mathbf{R}) -  \log\mathbf{b}
\end{equation}

where $\mathbf{b}$ is computed in a content-based manner by $ \mathbf{b} = -\mathrm{E}_{x \sim P(x)} [\mathbf{p}(y|x)] \approx -\frac{1}{M} \sum_{i=1}^M \mathbf{p}(y|x^{(i)})$ for prompt classification or  $ \mathbf{b} = -\mathrm{E}_{x,r \sim P(x,r)} [\mathbf{p}(y|x, r)] \approx -\frac{1}{M} \sum_{i=1}^M \mathbf{p}(y|x^{(i)}, r^{(i)})$ for response classification. Note that batch calibration requires a batch of unlabeled samples to estimate the contextual prior during test time.

\subsection{Calibration Results}

We apply the calibration methods discussed in Section~\ref{cal} to both prompt classification and response classification for each guard model. For temperature scaling, we utilize the XSTest set as the validation set to optimize the temperature due to its relatively small size. This optimized temperature value is then applied across all other datasets, as individual validation sets are not available for all examined datasets. Additional experiments using in-domain validation sets can be found in Appendix~\ref{app:in-domain-val}. For contextual calibration, we estimate the contextual bias using a space token. For batch calibration, we assume access to the full test set and estimate the contextual bias using the entire test set as a batch. The resulting calibration performance is reported in Table~\ref{table3}. Details regarding the implementation can be found in Appendix~\ref{implement_detail}, along with additional calibration results in adversarial environments in Appendix~\ref{more_cal}.

\input{table/table3}

\textbf{Contextual calibration proves more effective for prompt classification and temperature scaling benefits response classification more}. Empirical results indicate that contextual calibration outperforms other methods in prompt classification, delivering improved calibration for the majority of guard models, with the exception of WildGuard. Additionally, temperature scaling effectively reduces the ECE and demonstrates particular effectiveness, despite being optimized on a validation set with a potentially different distribution from the target dataset. This finding further confirms the shared overconfident predictions across datasets and validates that proper temperature values can smooth the overconfident prediction distribution, thereby mitigating miscalibration. Furthermore, temperature scaling shows greater efficacy in response classification which often involves multiple sentences of both user inputs and model responses. In such cases, contextual calibration struggles to accurately estimate contextual prior, resulting in unstable or even degraded calibration performance. Moreover, it is noteworthy that batch calibration underperforms compared to contextual calibration for most models in prompt classification, as well as some models in response classification. We conjecture that this could be attributed to significant label shifts in the test datasets, leading to biased contextual prior estimation and diminished calibration effectiveness. However, no single method fully resolves the miscalibration issues, indicating the complexity of achieving reliable safety moderation across different deployment scenarios.

\section{Discussion}
In this section, to further understand why miscalibration of guard models happens, and how it manifests in prompt and response classification, we conduct two further investigations and point out the limitation and weak robustness of instruction-tuning LLM-based guard models.

\input{table/table4_cf}

\textbf{Prediction Vulnerability Induced by a Single Token}. We analyze two specific scenarios by assessing model predictions when the user input consists of a space token or an ``unsafe'' token and both the user input and model response consist of a space token, respectively. Results of the probability of the input being classified as ``safe'' are reported in Table~\ref{table4}. The results demonstrate that many guard models exhibit high confidence in predicting ``safe'' for a space token input. However, the introduction of a single ``unsafe'' token without further context can cause many guard models to confidently predict ``unsafe''. This finding underscores the persistent contextual bias in guard models revealing their limitations even after instruction-tuning. More extensive robustness evaluations of guard models are thus essential for future research.

\textbf{Misaligned Classification Objectives}. We further investigate guard models in the LLama-Guard family capable of both prompt and response classification, focusing on the accuracy of predictions when the model response is set to a content-free token. Specifically, we sample 100 ``safe'' user inputs and 100 ``unsafe'' user inputs from the WildGuardTest set and replace all model responses with a space token. We report the average probability of classifying the response as ``safe'' for using ``safe'' and ``unsafe'' user inputs separately in Table~\ref{table5}. The results indicate that the model is more likely to predict the responses as ``unsafe'' when user inputs (prompt) are unsafe, even when model responses are content-free and should logically be predicted as ``safe''. This suggests that the model prediction is heavily influenced by the user input and the guard models act like conducting prompt classification even when response classification should be done. Such behavior can result in unreliable predictions and increased miscalibration.

\section{Conclusion}
In this work, we have systematically examined the uncertainty-based reliability of LLM-based guard models by assessing their calibration levels across various benchmarks. Our analysis reveals that despite their promising performance in content moderation, these models tend to make overconfident predictions, exhibit significant miscalibration under adversarial environments, and lack robustness to responses generated by diverse LLMs. To mitigate miscalibration, we explore several post-hoc calibration techniques. Our results show that contextual calibration proves particularly effective for prompt classification and temperature scaling improves response classification performance more. Our findings underscore the importance of uncertainty-based reliability and advocate for incorporating confidence calibration evaluation in the development and release of future LLM-based guard models. 

\section*{Ethical Statement and Broader Impacts}
Our work examines the reliability and confidence calibration of existing LLM-based guard models. Despite enhanced confidence calibration for certain guard models, it is essential to emphasize that confidence calibration should not be the sole criterion for determining the suitability of a guard model for deployment. Guard models could potentially make incorrect predictions, particularly when dealing with texts in the wild. We thus advocate for a more holistic reliability evaluation that integrates uncertainty-based confidence calibration with assessments of model robustness and overall performance, and in certain cases, even human-involved factors tailored to specific scenarios.   

The broader implications of our study have the potential to drive future research towards the development of better-calibrated guard models. Our insights also contribute to the design of more effective post-hoc calibration techniques, the incorporation of calibration optimization during instruction tuning, and the synthesis of diverse and high-quality data aimed at enhancing both calibration and robustness in the future. Furthermore, our work potentially provides valuable guidance on selecting classifiers that can ensure consistent and reliable evaluations of attack success rate (ASR) in determining whether a jailbreak attack has succeeded. 

\section*{Limitations}
Our work focuses on post-hoc calibration methods for open-source LLM-based guardrail models. The explored methods do not apply to closed-source models where the logit outputs are unavailable. As for each calibration method, there exist trade-offs. Temperature scaling requires the in-domain validation set for temperature optimization, but in-domain data are not always available in the practical setting. Contextual calibration requires access to the instruction prompt for inference, but the bias captured from content-free tokens may not always be accurate enough. Batch calibration requires access to a batch of unlabeled samples in the target domain, but they could be adapted to adversarial distribution shifts and may need additional validation sets to determine the batch size. In general, post-hoc calibration methods only mitigate the miscalibration in certain scenarios and it is challenging for one single method generalizable to all models and datasets. Nevertheless, this inspires future works to design not only better post-hoc calibration methods but also more reliable training methods to address miscalibration.   

\section*{Reproducibility Statement}
In support of reproducibility, our submission includes the full code implementation, along with comprehensive instructions in the README.md file detailing the steps required to install the necessary environments and run our experiments. Additionally, we provide all relevant information regarding publicly available datasets and models, enabling interested researchers can replicate the results presented in this paper. 

\section*{Acknowledgments}
The authors would like to thank anonymous reviewers for their valuable suggestions. This project is funded by a research grant MOE-MOESOL2021-0005 from the Ministry of Education in Singapore.

\newpage

\bibliography{iclr2025_conference}
\bibliographystyle{iclr2025_conference}

\newpage

\appendix
\section{Appendix}
\subsection{Dataset Details}
\label{dataset_stat}
In this section, we briefly describe the public datasets we examined and show the statistics in Table~\ref{table:app_data}.

\input{appendix/app_data1}

\textbf{OpenAI Moderation}~\citep{openai}. This dataset contains 1680 prompts with labels for 8 unsafe categories including sexual, hate, violence, harassment, self-harm, sexual/minors, hate/threatening, and violence/graphic. Each category label is a binary flag.

\textbf{ToxicChat Test}~\citep{toxicchat}. We use the test split of the new version toxicchat0124, involving 2853 user prompts collected from the Vicuna online demo\footnote{https://lmarena.ai/}, each annotated with binary toxicity labels.  

\textbf{Aegis Safety Test}~\citep{aegis}. This dataset is built on the prompts from HH-RLHF and responses generated by Mistral-7B-v0.1 with human annotations. We utilize the prompt-only subset, with a size of 359, from the test split of the dataset. The absence of the turn-level split of utterances during the conversation makes it infeasible for response classification evaluation. This dataset covers 13 unsafe content categories according to NVIDIA's content safety taxonomy including Hate/Identity Hate, Sexual, Violence, Suicide and Self Harm, Threat, Sexual Minor, Guns/Illegal Weapons, Controlled/Regulated substances, Criminal Planning/Confessions, PII, Harassment, Profanity, Other. The ``Needs Caution'' category is also involved for uncertain cases.   

\textbf{SimpleSafetyTests}~\citep{simplesafetytests}. This dataset involves 100 manually-crafted harmful prompts with topics in Suicide, Self-Harm and Eating Disorders, Physical Harm and Violence, Illegal and Highly Regulated items, Scams and Fraud, Child Abuse.

\textbf{XSTest}~\citep{xstest}. This dataset contains 250 safe prompts and 200 unsafe prompts. Safe prompts use similar language to unsafe prompts or mention sensitive topics but they are clearly safe and should not be refused. Binary labels are provided in this dataset.  

\textbf{Harmbench Prompt}~\citep{harmbench}. This dataset is designed for robustness to jailbreak attacks with prompts for eliciting harmful outputs from LLMs. We use the ``standard'' and ``copyright'' subsets, with a total size of 239, from the test split of the dataset in our evaluation for LLM-based guard models on prompt classification. The topics of unsafe prompts include Cybercrime \& Unauthorized Intrusion, Chemical \& Biological Weapons/Drugs, Copyright Violations, Misinformation \& Disinformation, Harassment \& Bullying, Illegal Activities, General Harm. 

\textbf{Harmbench Response}~\citep{harmbench}. This dataset refers to a variant of the validation set used for fine-tuning Llama2-variant from Harmbench, which consists of 602 responses generated by various models and jailbreak attacks. We use the pairs of their vanilla prompts and model responses with human labeling for response classification, resulting in a set of 596 pairs.   

\textbf{Harmbench-adv}~\citep{harmbench}. This dataset refers to the original validation set with a size of 602 for fine-tuning Llama2-variant from Harmbench. We term it ``Harmbench-adv'' to differentiate it from ``Harmbench Response'' given that adversarial prompts from diverse attack methods are involved in the Harmbench-adv set. Adversarial prompts could be very different from vanilla ones. We further split this dataset in terms of the type of response models and retain 10 subsets with statistics in Table~\ref{app:model_dep}.

\textbf{BeaverTail Test}~\citep{beavertails}. We utilize the test split of this dataset with 33.4k prompt-response pairs, which contain manually annotated labels for model response harmfulness. The prompts are modified from the HH-RLHF and Safety-Prompts, while the responses are generated with the Alpaca-7B model. The 14 harm categories involve Animal Abuse, Child Abuse, Controversial Topics \& Politics, Discrimination \& Stereotype \& Injustice, Drug Abuse \& Weapons \& Banned Substance, Financial Crime \& Property Crime \& Theft, Hate Speech \& Offensive Language, Misinformation Regarding ethics \& laws \& safety, Non-Violent Unethical Behavior, Privacy Violation, Self-Harm, Sexually Explicit \& Adult Content, Terrorism \& Organized Crime, Violence \& Aiding and Abetting \& Incitement. We use a subset of 2k size randomly sampled from the original test split to reduce the evaluation cost.

\textbf{SafeRLHF Test}~\citep{saferlhf}. The dataset shares the prompts with the BeaverTails dataset and generates responses from Alpaca-7B, Alpaca2-7B, and Alpaca3-8B. The 19 harm categories include Endangering National Security, Insulting Behavior, Discriminatory Behavior, Endangering Public Health, Copyright Issues, Violence, Drugs, Privacy Violation, Economic Crime, Mental Manipulation, Human Trafficking, Physical Harm, Sexual Content, Cybercrime, Disrupting Public Order, Environmental Damage, Psychological Harm, White-Collar Crime, Animal Abuse. We use a subset of 2k size randomly sampled from the original test split to reduce the evaluation cost.  

\textbf{WildGuardMix Test}~\citep{wildguard}. This dataset contains 1725 samples with synthetic, in-the-wild user-LLM interactions and annotator-written data. Responses to synthetic and vanilla prompts are generated using a suite of LLMs. We consider the prompt harmfulness and response harmfulness annotations in our evaluations. WildGuardMix Test Prompt and WildGuardMix Test Response refer to the prompts data and prompt+response pairs data for prompt and response classification, respectively.

\input{appendix/app_data2}

\subsection{Model Details}
\label{model_detail}
We summarize the hugging face model cards of 9 LLM-based guard models we examined in Table~\ref{app:guard_model_card}. Note that we do not assess the series of ShieldGemma\footnote{https://huggingface.co/google/shieldgemma-2b} models given that they only support classification for a single policy per inference, making public datasets infeasible for evaluation due to the policy difference. 

\input{appendix/app_model1}

\subsection{Implementation Details}
\label{implement_detail}
We use Pytorch and Huggingface Transformers in our implementation. We run all evaluations on a single NVIDIA A40 GPU (48G). We use $M=15$ bins as in \citet{guo2017calibration} for all our ECE evaluations. For temperature scaling, we optimize the $T$ within the range from (0, 5]. For batch calibration, we set the batch size as the size of the entire test set by default following ~\cite{zhou2023batch}. For prompt classification, we keep the original prompt lengths for most datasets except OpenAI Moderation where we truncate a few samples with extremely long lengths to avoid the out-of-memory error. We keep the maximum length as 1800. For response classification, we keep the original prompt length for all datasets and set the maximum response length as 500.

\subsection{More Reliability Diagrams}
\label{more_rediag}
\begin{figure}[h]
    \centering
    \includegraphics[width=\textwidth]{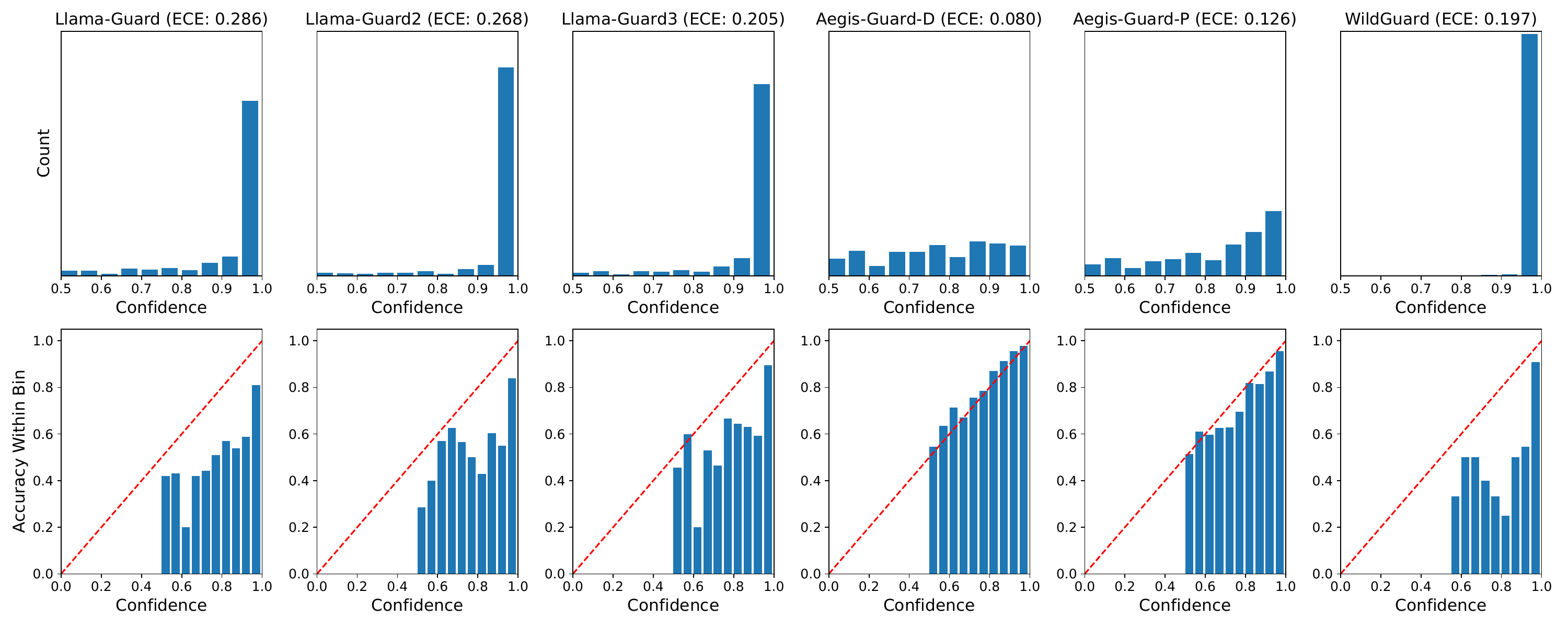}
    \caption{Confidence distributions (First row) and reliability diagrams (Second row) on the WildGuardMix Test Prompt set.}
    \label{fig:rediag_prompt_app1}
    \includegraphics[width=\textwidth]{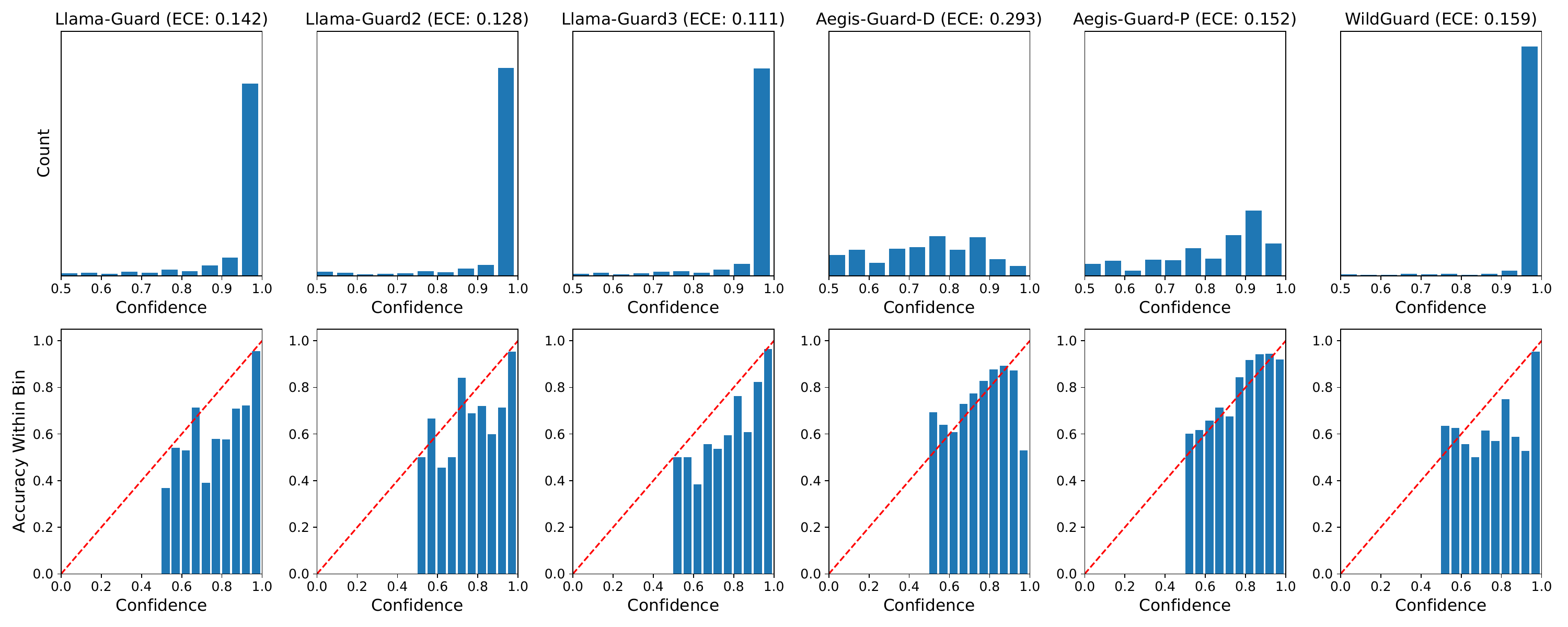}
    \caption{Confidence distributions (First row) and reliability diagrams (Second row) on the WildGuardMix Test Response set.}
    \label{fig:rediag_prompt_res_app1}
    \includegraphics[width=0.5\textwidth]{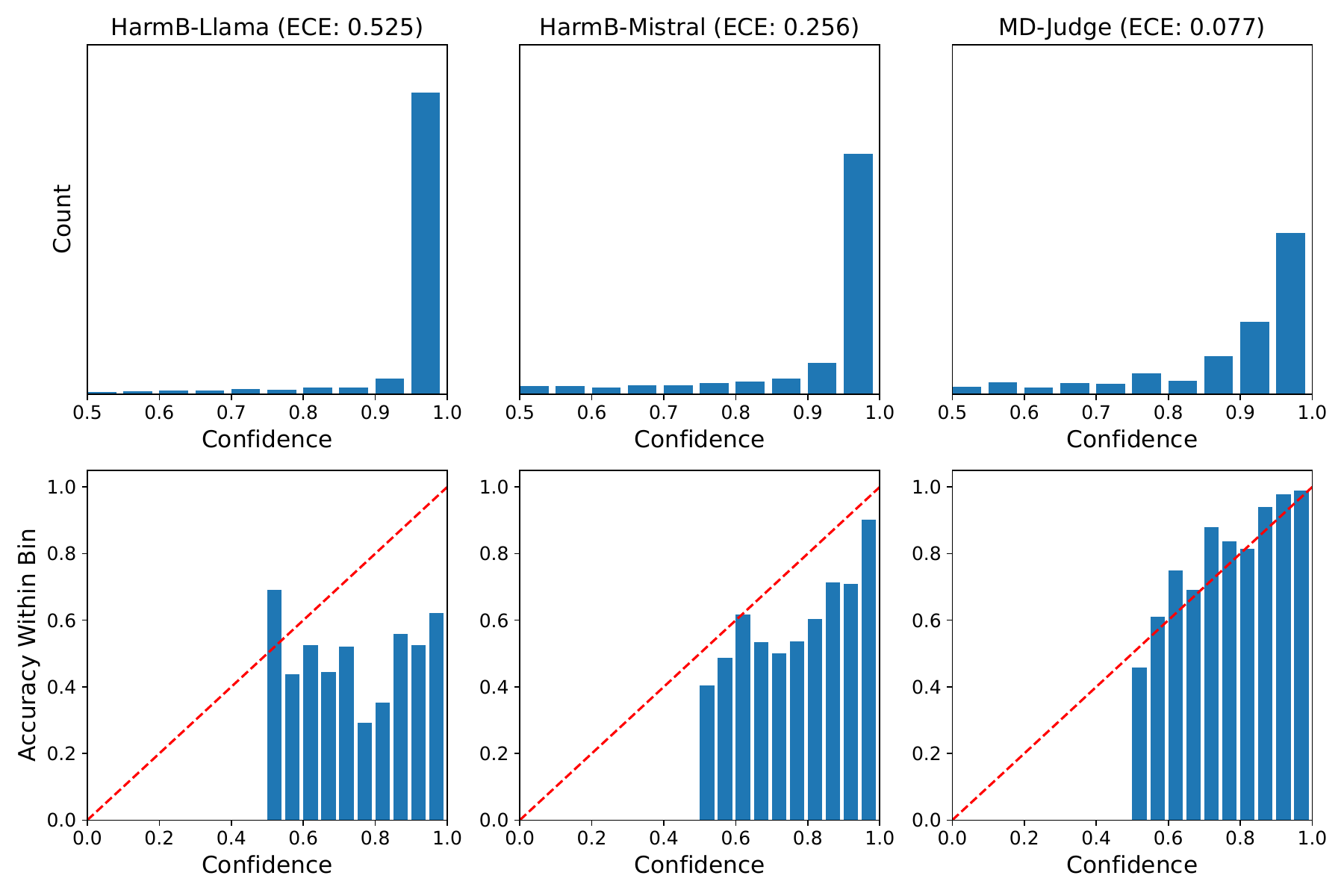}
    \caption{Confidence distributions (First row) and reliability diagrams (Second row) on the WildGuardMix Test Response set.}
    \label{fig:rediag_prompt_res_app2}
\end{figure}

\begin{figure}[h]
    \centering
    \includegraphics[width=\textwidth]{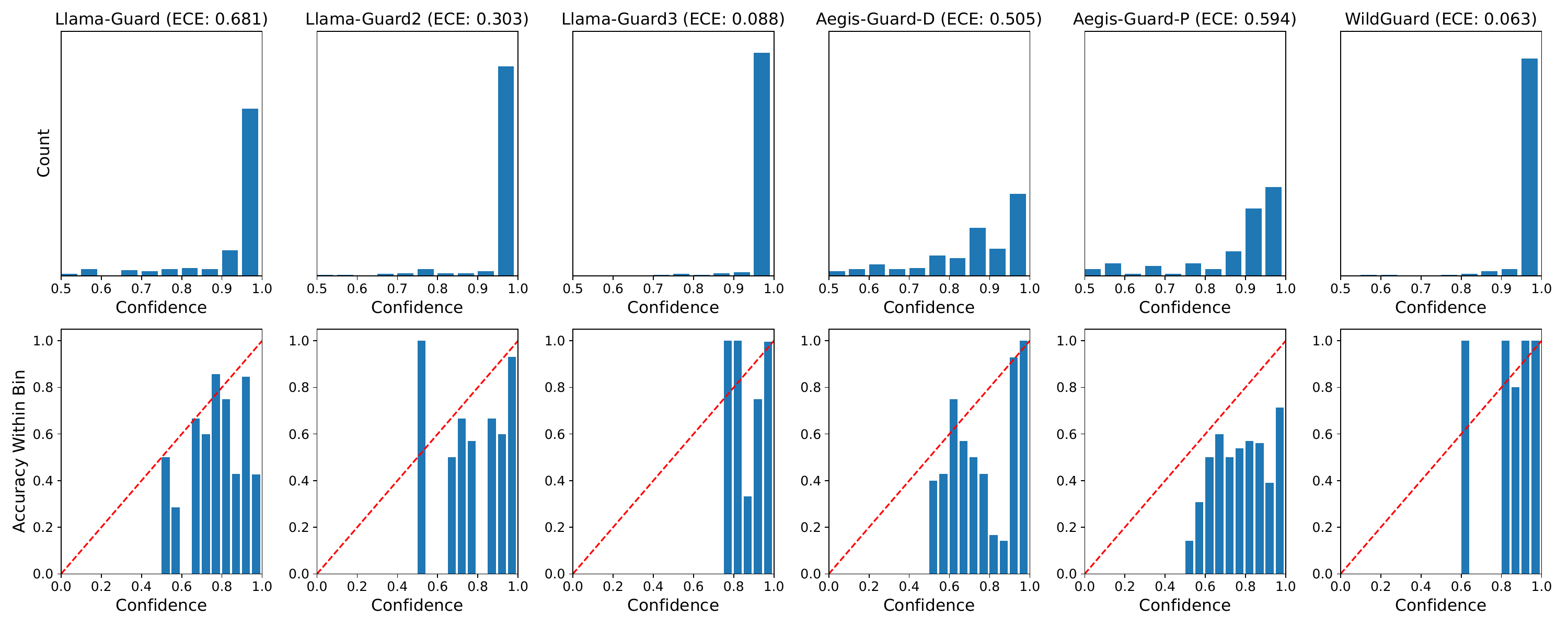}
    \caption{Confidence distributions (First row) and reliability diagrams (Second row) on the Harmbench Prompt set.}
    \label{fig:rediag_prompt_harm_app1}
    \includegraphics[width=\textwidth]{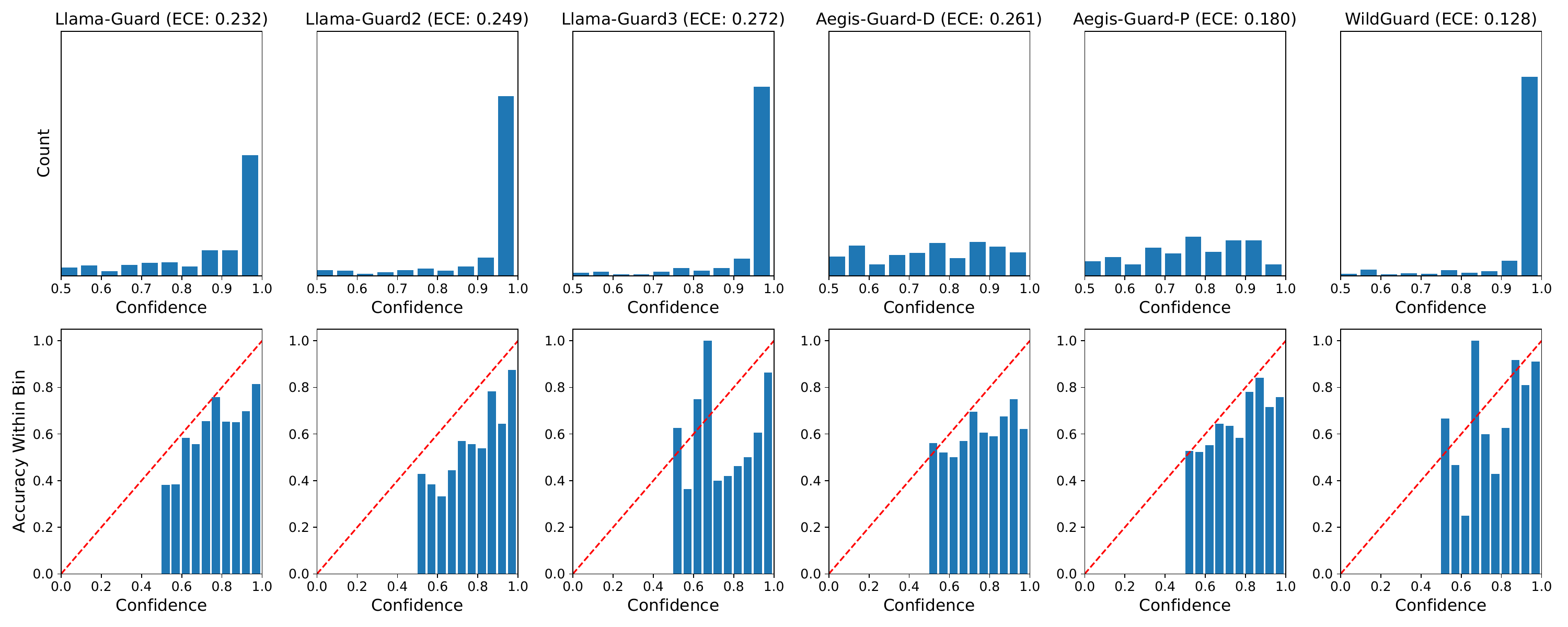}
    \caption{Confidence distributions (First row) and reliability diagrams (Second row) on the Harmbench Response set.}
    \label{fig:rediag_prompt_harm_res_app1}
    \includegraphics[width=0.5\textwidth]{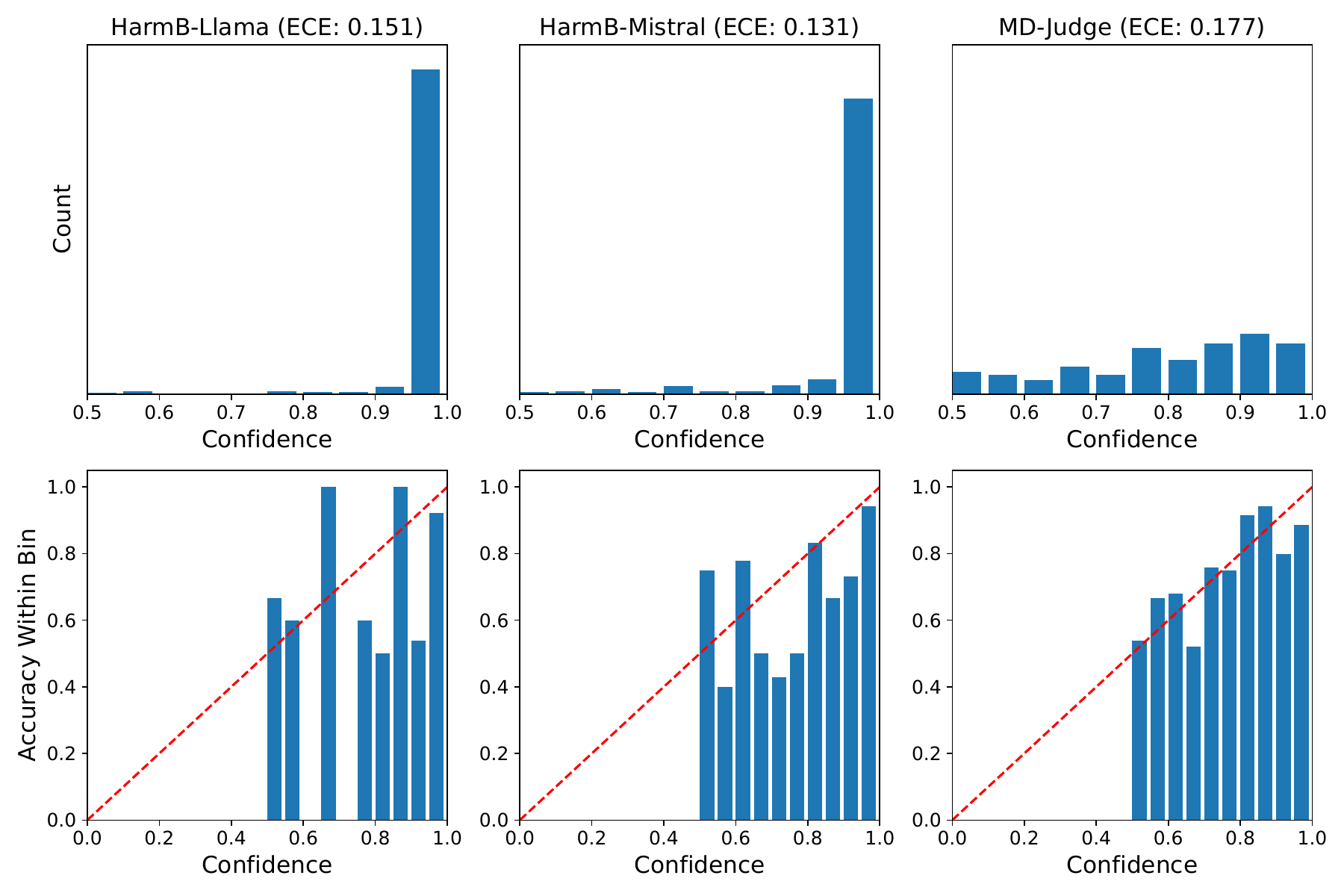}
    \caption{Confidence distributions (First row) and reliability diagrams (Second row) on the Harmbench Response set.}
    \label{fig:rediag_prompt_harm_res_app2}
\end{figure}

We report the full reliability diagrams with corresponding confidence distributions of all 9 models we examined for both prompt classification and response classification on the WildGuardMix Test Prompt set, shown in Figure~\ref{fig:rediag_prompt_app1}, the WildGuardMix Test Response set, shown in Figure~\ref{fig:rediag_prompt_res_app1}, \ref{fig:rediag_prompt_res_app2}, the Harmbench Prompt set, shown in Figure~\ref{fig:rediag_prompt_harm_app1}, and the Harmbench Response set, shown in Figure~\ref{fig:rediag_prompt_harm_res_app1}, \ref{fig:rediag_prompt_harm_res_app2}. The full results illustrate that existing LLM-based guard models exhibit varying levels of miscalibration. Models including LLama-Guard, LLama-Guard2, LLama-Guard3, WildGuard, Harmbench-Llama, and Harmbench-Mistral tend to make overconfident predictions, while Aegis-Guard-D, Aegis-Guard, and MD-Judge-v0.1 are not.  

\subsection{More Calibration Results under Adversarial Conditions}
\label{more_cal}
We conduct more calibration experiments using temperature scaling, contextual calibration, and batch calibration on the Harmbench-adv set to mitigate the miscalibration discussed in Section~\ref{eval_jail}, and Section~\ref{diverse_model}. We keep the same settings and implementations as those in the main text. The ECE results are presented in Table~\ref{app:adv_results} and Table~\ref{app:adv_model_dep}, respectively. These additional empirical findings indicate the same conclusion as in the main text, that contextual calibration proves impressively effective on prompt classification, while temperature scaling benefits more on response classification.  

\input{appendix/app_adv_results}

\input{appendix/app_adv_model_dep}

\subsection{Instruction Prompts}
Instruction prompts for all LLM-based guard models we examined can be found in our submitted code implementation or their Huggingface model cards in Table~\ref{app:guard_model_card}.

\clearpage

\section{Additional Experiments}

\subsection{Effects of in-domain validation sets for temperature scaling}
\label{app:in-domain-val}
To further improve the calibration effect of temperature scaling, we conduct additional experiments by optimizing temperature on the in-domain set. Specifically, we consider the WildGuardMix dataset and split a validation set of size 100 from its training set as the in-domain validation set. The temperature is then optimized on the new validation set. The results for prompt and response classification are reported in Table~\ref{app:ts_valid}. It is observed that the temperature optimized on the in-domain validation set is more effective in reducing ECE than the one optimized on XSTest. Despite the benefits, there is not always access to such an in-domain validation dataset for different scenarios in the wild, leading to a trade-off when applying temperature scaling.     

\input{appendix/app_ts_valid}

\subsection{Ablation on the length of adversarial prompts}

To investigate whether the guard model is vulnerable due to the jailbroken nature of the prompts, or due to some spurious correlations such as length, we design additional experiments. Specifically, we consider three different ranges of adversarial prompt length, leading to three subsets. Then we assess the ECE on three subsets separately and report the results in Table~\ref{app:abl_length}. It is observed that there are similar ECE performances among different length ranges, which further supports that it is the jailbroken nature of adversarial prompts that makes guard models vulnerable. We conjecture that there exist many unseen adversarial prompts with any token combinations in the wild and it is impossible to involve all adversarial prompts in instruction tuning of guard models. Thus, the evaluation of adversarial prompts introduces varying levels of uncertainty instead of making overconfident predictions that lead to high ECE performances.

\input{appendix/app_abl_length}

\subsection{Robustness on black-box response models}
To further assess the ECE performance when classifying responses from proprietary models, we try to collect the responses from GPT-3.5, GPT-4, and Claude-2 to the same set of adversarial queries, and then conduct the same evaluation of ECE. The results are reported in Table~\ref{app:model_comparison}. It is observed that all guard models have low ECE values when classifying responses from GPT-4 and Claude-2. The reason is that the utilized adversarial attacks are not effective enough for these well-aligned black-box models and thus elicit clear refusal responses that are easy to classify for guard models. Nevertheless, some guard models still exhibit high ECE values when classifying responses from GPT-3.5, serving as a nice complement and support to our Finding 3 in Section~\ref{diverse_model}. 

\input{appendix/app_blackbox_response_model}

\newpage
\subsection{Comparison of False Positive rate and false negative rates}
To investigate what types of over-confidence behaviors different guard models are showing, we report the statistics of the false positive rates (FPR) and false negative rates (FNR) in Table~\ref{app:fpr_fnr}. It is observed that most models generally showcase higher false negative rates, suggesting their over-confidence behaviors to predict the input as the safe type.  

\input{appendix/app_fpr_fnr}

\end{document}

%% file: math_commands.tex
\usepackage{amsmath,amsfonts,bm}

\def\eqref#1{equation~\ref{#1}}

\def\1{\bm{1}}

\DeclareMathAlphabet{\mathsfit}{\encodingdefault}{\sfdefault}{m}{sl}
\SetMathAlphabet{\mathsfit}{bold}{\encodingdefault}{\sfdefault}{bx}{n}



%% file: table/table1.tex
\begin{table}[htbp]
\centering

\begin{adjustbox}{max width=\textwidth}
\begin{tabular}{lcccccccc|cccccc}
\toprule[1pt]
\multirow{2}{*}{\textbf{Model}} & \multicolumn{8}{c|}{\textbf{Prompt Classification}} & \multicolumn{5}{c}{\textbf{Response Classification}} \\ 
               & \textbf{OAI} & \textbf{ToxiC} & \textbf{SimpST} & \textbf{Aegis} & \textbf{XST} & \textbf{HarmB} & \textbf{WildGT} & \textbf{Avg.} & \textbf{BeaverT} & \textbf{S-RLHF} & \textbf{HarmB} & \textbf{WildGT} & \textbf{Avg.} \\ 
\midrule[0.5pt]
Llama-Guard    & 9.0  & 11.0 & 32.6 & 29.6 & 20.5 & 68.1 & 28.6 & 28.5 & 29.1 & 24.4 & 23.2 & 14.2 & 22.7 \\
Llama-Guard2   & 13.7 & 15.9 & 26.5 & 34.7 & 12.2 & 30.3 & 26.8 & 22.9 & 29.9 & 25.2 & 24.9 & 12.8 & 23.2 \\
Llama-Guard3   & 13.9 & 14.5 & 7.6  & 33.4 & 13.9 & 8.8  & 20.5 & \underline{16.1} & 32.3 & 25.3 & 27.2 & 11.1 & 24.0 \\
Aegis-Guard-D  & 30.2 & 20.5 & 9.7  & 16.4 & 23.5 & 50.5 & 8.0  & 22.7 & 18.1 & 30.9 & 26.1 & 29.3 & 26.1 \\
Aegis-Guard-P  & 15.7 & 8.2  & 16.5 & 22.6 & 18.6 & 59.4 & 12.6 & 22.0 & 18.6 & 25.6 & 18.0 & 15.2 & 19.4 \\
HarmB-Llama    & -    & -    & -    & -    & -   & - & -    & - & 24.9 & 19.4 & 15.1 & 52.5 & 28.0 \\
HarmB-Mistral  & -    & -    & -    & -    & -   & - & -    & - & 18.1 & 14.5 & 13.1 & 25.6 & \underline{17.8} \\
MD-Judge       & -    & -    & -    & -    & -   & - & -    & - & 10.9 & 9.4  & 17.7 & 7.7  & \textbf{11.4} \\
WildGuard      & 33.8 & 19.8 & 4.4  & 12.0 & 5.0  & 6.3 & 19.7 & \textbf{14.4} & 23.2 & 23.3 & 12.8 & 15.9 & 18.8 \\ 
\bottomrule[1pt]
\end{tabular}
\end{adjustbox}
\caption{ECE (\%) $\downarrow$ performances of prompt and response classification on existing public benchmarks. We bold the best average result and underline the second-best average result for both prompt and response classification.}
\label{table1}
\end{table}

%% file: table/table2.tex
\begin{table}[htbp]
\centering
\begin{adjustbox}{max width=\textwidth}
\begin{tabular}{lc|cccccccccc}
\toprule[1pt]
\multirow{2}{*}{\textbf{Guard Model}} & \multicolumn{11}{c}{\textbf{Response Model}}  \\ 
               & \textbf{Metric} & \textbf{Baichuan2} & \textbf{Qwen} & \textbf{Solar} & \textbf{Llama2} & \textbf{Vicuna} & \textbf{Orca2} & \textbf{Koala} & \textbf{OpenChat} & \textbf{Starling} & \textbf{Zephyr} \\ 
\midrule[0.5pt]
\multirow{2}{*}{Llama-Guard} & F1 & 57.8 & 66.7 & 54.2 & 44.4 & 64.0 & 62.7 & 74.6 & 60.6 & 66.7 & 72.7 \\
& ECE & 26.9 & 23.0 & 49.4 & 10.5 & 28.0 & 26.4 & 27.4 & 40.3 & 46.3 & 38.5 \\
\midrule[0.5pt]
\multirow{2}{*}{Llama-Guard2} & F1 & 77.8 & 88.9 & 82.8 & 71.4 & 72.1 & 80.6 & 78.4 & 70.0 & 82.0 & 78.4 \\
& ECE & 18.2 & 5.8 & 27.1 & 7.9 & 28.4 & 25.2 & 30.4 & 28.5 & 37.0 & 39.4 \\
\midrule[0.5pt]
\multirow{2}{*}{Llama-Guard3} & F1 & 73.8 & 82.4 & 84.1 & 60.0 & 83.3 & 82.2 & 77.5 & 76.4 & 91.2 & 87.7 \\
& ECE & 33.7 & 17.1 & 31.0 & 27.4 & 20.5 & 27.3 & 36.5 & 34.2 & 27.6 & 23.1 \\
\midrule[0.5pt]
\multirow{2}{*}{Aegis-Guard-D} & F1 & 60.3 & 66.7 & 71.2 & 31.2 & 63.3 & 65.8 & 69.9 & 78.3 & 84.4 & 89.3 \\
& ECE & 35.5 & 27.1 & 22.2 & 40.8 & 34.0 & 33.9 & 31.3 & 30.9 & 27.8 & 30.9 \\
\midrule[0.5pt]
\multirow{2}{*}{Aegis-Guard-P} & F1 & 57.6 & 66.7 & 67.9 & 33.3 & 56.3 & 72.7 & 72.2 & 76.2 & 83.3 & 80.8 \\
& ECE & 22.8 & 17.4 & 28.3 & 23.6 & 26.2 & 28.2 & 32.1 & 35.5 & 36.3 & 25.7 \\
\midrule[0.5pt]
\multirow{2}{*}{HarmB-Llama} & F1 & 89.7 & 100.0 & 90.6 & 70.6 & 90.9 & 86.2 & 88.9 & 89.4 & 90.9 & 94.5 \\
& ECE & 17.7 & 6.4 & 25.0 & 23.5 & 16.0 & 19.5 & 26.7 & 23.2 & 23.3 & 20.1 \\
\midrule[0.5pt]
\multirow{2}{*}{HarmB-Mistral} & F1 & 84.4 & 100.0 & 87.5 & 80.0 & 92.3 & 84.8 & 92.8 & 90.9 & 89.2 & 94.5 \\
& ECE & 28.0 & 3.0 & 30.1 & 12.8 & 16.6 & 17.5 & 16.0 & 14.9 & 27.7 & 19.9 \\
\midrule[0.5pt]
\multirow{2}{*}{MD-Judge} & F1 & 75.4 & 79.1 & 77.2 & 55.6 & 74.2 & 76.9 & 75.3 & 76.6 & 87.5 & 92.6 \\
& ECE & 22.4 & 14.4 & 19.3 & 24.1 & 19.9 & 16.7 & 26.2 & 25.5 & 26.0 & 17.9 \\
\midrule[0.5pt]
\multirow{2}{*}{WildGuard} & F1 & 82.0 & 91.3 & 88.5 & 80.0 & 89.9 & 84.8 & 81.6 & 88.9 & 92.5 & 94.5 \\
& ECE & 22.1 & 9.2 & 15.5 & 17.0 & 11.2 & 20.1 & 37.3 & 25.4 & 18.6 & 21.3 \\
\bottomrule[1pt]
\end{tabular}
\end{adjustbox}
\caption{F1 (\%) $\uparrow$ and ECE (\%) $\downarrow$ performances of response classification on Harmbench-adv set across 10 different response models.}
\label{model_dep}
\end{table}

%% file: table/table3.tex
\begin{table}[htbp]
\centering

\begin{adjustbox}{max width=\textwidth}
\begin{tabular}{lcccccccc|cccccc}
\toprule[1pt]
\multirow{2}{*}{\textbf{Model}} & \multicolumn{8}{c|}{\textbf{Prompt Classification}} & \multicolumn{5}{c}{\textbf{Response Classification}} \\ 
               & \textbf{OAI} & \textbf{ToxiC} & \textbf{SimpST} & \textbf{Aegis} & \textbf{XST} & \textbf{HarmB} & \textbf{WildGT} & \textbf{Avg.} & \textbf{BeaverT} & \textbf{S-RLHF} & \textbf{HarmB} & \textbf{WildGT} & \textbf{Avg.} \\ 
\midrule[0.5pt]
Llama-Guard    & 9.0  & 11.0 & 32.6 & 29.6 & 20.5 & 68.1 & 28.6 & 28.5 & 29.1 & 24.4 & 23.2 & 14.2 & 22.7 \\
+ TS           & 12.0 & 11.3 & 31.9 & 26.8 & 9.4 & 66.7 & 26.0 & 26.3 & 27.4 & 21.6 & 14.5 & 14.0 & 19.4 \\
+ CC           & 14.8 & 7.4 & 26.3 & 22.0 & 23.9 & 65.1 & 20.9 & \textbf{25.8} & 25.4 & 21.8 & 20.2 & 8.9 & \textbf{19.1} \\
+ BC           & 12.3 & 12.1 & 43.2 & 27.2 & 21.0 & 67.9 & 19.7 & 29.1 & 26.6 & 22.5 & 20.6 & 12.4 & 20.5 \\
\midrule[0.5pt]
Llama-Guard2   & 13.7 & 15.9 & 26.5 & 34.7 & 12.2 & 30.3 & 26.8 & 22.9 & 29.9 & 25.2 & 24.9 & 12.8 & 23.2 \\
+ TS   & 13.2 & 15.8 & 26.0 & 33.6 & 11.1 & 29.4 & 26.0 & 22.2 & 29.8 & 24.5 & 14.1 & 13.6 & \textbf{20.5} \\
+ CC   & 39.4 & 22.8 & 15.0 & 18.8 & 13.7 & 14.8 & 15.3 & \textbf{20.0} & 24.3 & 28.9 & 34.8 & 32.8 & 30.2 \\
+ BC   & 15.2 & 16.6 & 30.6 & 34.2 & 12.0 & 36.3 & 23.8 & 24.1 & 29.5 & 25.2 & 24.8 & 14.7 & 23.6 \\
\midrule[0.5pt]
Llama-Guard3   & 13.9 & 14.5 & 7.6  & 33.4 & 13.9 & 8.8 & 20.5 & 16.1 & 32.3 & 25.3 & 27.2 & 11.1 & 24.0 \\
+ TS   & 16.0 & 15.2 & 9.3 & 30.6 & 11.5 & 9.8 & 20.4 & 16.1 & 31.2 & 24.8 & 19.3 & 13.0 & \textbf{22.0} \\
+ CC   & 28.3 & 21.7 & 4.1 & 26.3 & 8.0 & 4.5 & 15.0  & \textbf{15.4} & 21.9 & 30.6 & 39.7 & 30.3 & 30.6 \\
+ BC   & 17.1 & 20.5 & 24.9 & 32.2 & 13.0 & 23.3 & 18.1 & 21.3 & 30.5 & 24.7 & 25.7 & 14.0 & 23.7 \\
\midrule[0.5pt]
Aegis-Guard-D  & 30.2 & 20.5 & 9.7  & 16.4 & 23.5 & 50.5 & 8.0 & 22.7 & 18.1 & 30.9 & 26.1 & 29.3 & 26.1 \\
+ TS  & 30.1 & 25.2 & 14.4 & 11.1 & 18.7 & 46.9 & 13.4 & 22.8 & 15.7 & 25.9 & 23.8 & 31.3 & 24.2 \\
+ CC  & 15.6 & 9.5 & 17.7 & 22.0 & 16.3 & 58.7 & 12.4 & \textbf{21.7} & 17.8 & 24.3 & 17.5 & 16.3 & \textbf{19.0} \\
+ BC  & 24.9 & 34.6 & 43.0 & 18.4 & 20.5 & 56.2 & 9.0 & 29.5 & 17.2 & 23.1 & 18.7 & 34.2 & 23.3 \\
\midrule[0.5pt]
Aegis-Guard-P  & 15.7 & 8.2  & 16.5 & 22.6 & 18.6 & 59.4 & 12.6 & 22.0 & 18.6 & 25.6 & 18.0 & 15.2 & 19.4 \\
+ TS  & 19.7 & 14.9 & 20.0 & 16.7 & 11.4 & 55.6 & 13.8 & \textbf{21.7} & 16.5 & 21.6 & 18.2 & 19.3 & \textbf{18.9} \\
+ CC  & 17.6 & 9.2 & 15.3 & 21.5 & 19.3 & 58.5 & 11.1 & 21.8 & 18.3 & 26.3 & 18.7 & 16.6 & 20.0 \\
+ BC  & 19.7 & 27.9 & 43.5 & 22.7 & 18.8 & 61.1 & 6.9 & 28.7 & 18.6 & 23.5 & 17.6 & 28.5 & 22.1 \\
\midrule[0.5pt]
HarmB-Llama    & -    & -    & -    & -    & -   & - & -    & - & 24.9 & 19.4 & 15.1 & 52.5 & 28.0 \\
+ TS    & -    & -    & -    & -    & -   & - & -    & - & 22.2 & 17.3 & 14.2 & 51.1 & \textbf{26.2} \\
+ CC    & -    & -    & -    & -    & -   & - & -    & - & 34.9 & 32.0 & 24.9 & 61.7 & 38.4 \\
+ BC    & -    & -    & -    & -    & -   & - & -    & - & 24.2 & 18.8 & 14.9 & 52.3 & 27.6 \\
\midrule[0.5pt]
HarmB-Mistral  & -    & -    & -    & -    & -   & - & -    & - & 18.1 & 14.5 & 13.1 & 25.6 & 17.8 \\
+ TS  & -    & -    & -    & -    & -   & - & -    & - & 13.4 & 10.3 & 11.1 & 23.4 & \textbf{14.6} \\
+ CC  & -    & -    & -    & -    & -   & - & -    & - & 19.9 & 23.5 & 22.4 & 40.2 & 26.5 \\
+ BC  & -    & -    & -    & -    & -   & - & -    & - & 18.2 & 14.5 & 12.7 & 30.7 & 19.0 \\
\midrule[0.5pt]
MD-Judge       & -    & -    & -    & -    & -   & - & -    & - & 10.9 & 9.4  & 17.7 & 7.7  & \textbf{11.4} \\
+ TS       & -    & -    & -    & -    & -   & - & -    & - & 12.7 & 12.1 & 20.1 & 13.0 & 14.5 \\
+ CC       & -    & -    & -    & -    & -   & - & -    & - & 22.1 & 35.9 & 41.3 & 33.7 & 33.3 \\
+ BC       & -    & -    & -    & -    & -   & - & -    & - & 9.9 & 8.3 & 17.1 & 22.9 & 14.6 \\
\midrule[0.5pt]
WildGuard      & 33.8 & 19.8 & 4.4  & 12.0 & 5.0  & 6.3 & 19.7 & 14.4 & 23.2 & 23.3 & 12.8 & 15.9 & 18.8 \\ 
+ TS      & 32.4 & 19.1 & 5.7 & 9.1 & 4.2 & 8.2 & 19.3 & \textbf{14.0} & 23.8 & 22.3 & 10.5 & 16.5 & \textbf{18.3} \\ 
+ CC      & 58.7 & 39.0 & 0.2 & 26.5 & 25.5 & 0.1 & 18.6 & 24.1 & 22.8 & 27.9 & 16.2 & 16.1 & 20.8 \\ 
+ BC      & 33.6 & 23.8 & 25.2 & 12.7 & 3.8 & 30.6 & 19.5 & 21.3 & 23.1 & 22.2 & 12.6 & 16.3 & 18.6 \\ 
\bottomrule[1pt]
\end{tabular}
\end{adjustbox}
\caption{ ECE (\%) $\downarrow$ performance comparison of different calibration techniques. For each guard model, we report the original calibration results in the first row and the rest results using TS: Temperature Scaling, CC: Contextual Calibration, BC: Batch Calibration, in the following three rows. We bold the best average result among different calibration techniques for each guard model in both prompt and response classification.}
\label{table3}
\end{table}

%% file: table/table4_cf.tex
\begin{wraptable}{r}{0.5\textwidth}
\centering
\vskip -.13in
\begin{adjustbox}{max width=0.5\textwidth}
\begin{tabular}{lcc|c}
\toprule[1pt]
\multirow{2}{*}{\textbf{Model}} & \multicolumn{2}{c|}{\textbf{Prompt }} & \multicolumn{1}{c}{\textbf{Response }} \\ 
               & \textbf{Space} & \textbf{``unsafe''} & \textbf{Space} \\ 
\midrule[0.5pt]
Llama-Guard    & 75.5 & 18.2 & 73.1  \\
Llama-Guard2   & 98.9 & 83.5 & 99.2  \\
Llama-Guard3   & 90.5 & 53.1 & 98.8  \\
Aegis-Guard-D  & 29.4 & 9.5  & 29.4  \\
Aegis-Guard-P  & 53.1 & 16.5 & 53.1  \\
HarmB-Llama    & -    & -    & 98.8  \\
HarmB-Mistral  & -    & -    & 91.7  \\
MD-Judge       & -    & -    & 89.3  \\
WildGuard      & 99.5 & 92.0 & 77.7  \\ 
\bottomrule[1pt]
\end{tabular}
\end{adjustbox}
\caption{Probability (\%) of ``safe'' for prompt classification when input is set as a space token or ``unsafe'' token, and response classification when input and model output are set as a space token.}
\label{table4}
\vskip +.1in

\begin{adjustbox}{max width=0.5\textwidth}
\begin{tabular}{lcc}
\toprule[1pt]
\textbf{Model} & \textbf{Safe prompt} & \textbf{Unsafe prompt}  \\ 
\midrule[0.5pt]
Llama-Guard    & 62.1 & 29.0  \\
Llama-Guard2   & 63.0 & 21.5  \\
Llama-Guard3   & 62.9 & 16.4  \\
\bottomrule[1pt]
\end{tabular}
\end{adjustbox}
\caption{Probability (\%) of ``safe'' for response classification when output is set as a space token and inputs are sampled from safe/unsafe prompts. }
\label{table5}
\vskip -.2in

\end{wraptable}

%% file: appendix/app_data1.tex
\begin{table}[htbp]
\begin{center}
\begin{tabular}{lcc}
\toprule[1pt]
\multirow{2}{*}{\textbf{Dataset}} & \# \textbf{Prompt} & \# \textbf{Response}  \\ 
& safe/unsafe & safe/unsafe \\
\midrule[0.5pt]
Prompt Classification & & \\
\midrule[0.5pt]
OpenAI Moderation & 1158/522 & - \\
ToxicChat Test & 2491/362 & - \\
Aegis Safety Test & 126/233 & - \\
SimpleSafetyTests & 0/100 & - \\
XSTest & 250/200 & - \\
Harmbench Prompt & 0/239 & - \\
WildGuardMix Test Prompt & 971/754 & - \\
\midrule[0.5pt]
Response Classification & & \\
\midrule[0.5pt]
BeaverTails & - & 894/1106 \\
SafeRLHF & - & 1000/1000 \\
Harmbench Response & - & 326/270 \\
WildGuardMix Test Response & - & 1440/285 \\
\bottomrule[1pt]
\end{tabular}
\caption{Statistics of datasets we used.}
\label{table:app_data}
\end{center}
\end{table}

%% file: appendix/app_data2.tex
\begin{table}[htbp]
\centering
\begin{adjustbox}{max width=\textwidth}
\begin{tabular}{lc|cccccccccc}
\toprule[1pt]
\multirow{2}{*}{} & \multicolumn{11}{c}{\textbf{Response Model}}  \\ 
               & \textbf{Total} & \textbf{Baichuan2} & \textbf{Qwen} & \textbf{Solar} & \textbf{Llama2} & \textbf{Vicuna} & \textbf{Orca2} & \textbf{Koala} & \textbf{OpenChat} & \textbf{Starling} & \textbf{Zephyr} \\ 
\midrule[0.5pt]
\# \textbf{Response} & 540 & 64 & 62 & 45 & 69 & 68 & 65 & 59 & 38 & 37 & 33 \\
\bottomrule[1pt]
\end{tabular}
\end{adjustbox}
\caption{Statistics of subsets across 10 different response models.}
\label{app:model_dep}
\end{table}

%% file: appendix/app_model1.tex
\begin{table}[ht]
    \centering
    \begin{tabular}{l|l}
        \toprule
        \textbf{Guard Models}          & \textbf{Hugging Face page} \\
        \midrule
        Llama-Guard-7B          & \href{https://huggingface.co/meta-llama/LlamaGuard-7b}{meta-llama/LlamaGuard-7b} \\
        Llama-Guard2-8B            & \href{https://huggingface.co/meta-llama/Meta-Llama-Guard-2-8B}{meta-llama/Meta-Llama-Guard-2-8B} \\
        Llama-Guard3-8B        & \href{https://huggingface.co/meta-llama/Llama-Guard-3-8B}{meta-llama/Llama-Guard-3-8B} \\
        Aegis-Guard-Defensive-7B          & \href{https://huggingface.co/nvidia/Aegis-AI-Content-Safety-LlamaGuard-Defensive-1.0}{nvidia/Aegis-AI-Content-Safety-LlamaGuard-Defensive-1.0} \\
        Aegis-Guard-Permissive-7B          & \href{https://huggingface.co/nvidia/Aegis-AI-Content-Safety-LlamaGuard-Permissive-1.0}{nvidia/Aegis-AI-Content-Safety-LlamaGuard-Permissive-1.0} \\
        Harmbench-Llama2-13B          & \href{https://huggingface.co/cais/HarmBench-Llama-2-13b-cls}{cais/HarmBench-Llama-2-13b-cls} \\
        Harmbench-Mistral-7B          & \href{https://huggingface.co/cais/HarmBench-Mistral-7b-val-cls}{cais/HarmBench-Mistral-7b-val-cls} \\
        MD-Judge-v0.1-7B          & \href{https://huggingface.co/OpenSafetyLab/MD-Judge-v0.1}{OpenSafetyLab/MD-Judge-v0.1} \\
        WildGuard-7B          & \href{https://huggingface.co/allenai/wildguard}{allenai/wildguard} \\
        \bottomrule
    \end{tabular}
    \caption{Hugging Face Model Cards for examined LLM-based guard models}
    \label{app:guard_model_card}
\end{table}

%% file: appendix/app_adv_results.tex
\begin{table}[htbp]
\centering

\begin{adjustbox}{max width=\textwidth}
\begin{tabular}{lcccc|cccc}
\toprule[1pt]
\multirow{2}{*}{\textbf{Model}} & \multicolumn{4}{c|}{\textbf{Prompt Classification}} & \multicolumn{4}{c}{\textbf{Response Classification}} \\ 
               & \textbf{Origin} & \textbf{TS} & \textbf{CC} & \textbf{BC} & \textbf{Origin} & \textbf{TS} & \textbf{CC} & \textbf{BC} \\ 
\midrule[0.5pt]
Llama-Guard    & 69.3 & 65.6 & 58.7 & 65.8 & 30.8 & 30.1 & 21.2 & 19.0 \\
Llama-Guard2   & 59.6 & 57.6 & 35.0 & 61.2 & 24.4 & 13.7 & 31.1 & 23.7 \\
Llama-Guard3   & 46.2 & 44.5 & 36.3 & 50.7 & 20.8 & 8.8 & 33.2 & 20.8 \\
Aegis-Guard-D  & 40.4 & 40.8 & 54.5 & 52.2 & 21.5 & 21.8 & 11.8 & 13.8 \\
Aegis-Guard-P  & 54.4 & 51.9 & 52.5 & 56.6 & 12.1 & 16.7 & 12.5 & 12.5 \\ 
HarmB-Llama    & - & - & - & - & 27.1 & 26.3 & 35.2 & 26.3 \\ 
HarmB-Mistral  & - & - & - & - & 19.8 & 17.9 & 29.1 & 18.7 \\ 
MD-Judge       & - & - & - & - & 12.9 & 17.2 & 32.9 & 12.9  \\ 
WildGuard      & 34.9 & 34.3 & 26.2 & 38.6 & 13.1 & 10.4 & 14.3 & 13.1 \\  
\midrule[0.5pt]
Avg.           & 54.0 & 52.1 & \textbf{43.3} & 57.3 & 20.3 & \textbf{17.1} & 24.6 & 17.9 \\  
\bottomrule[1pt]
\end{tabular}
\end{adjustbox}
\caption{ ECE (\%) $\downarrow$ performance comparison of different calibration techniques. For each guard model, we report the Origin: original results, TS: Temperature Scaling, CC: Contextual Calibration, BC: Batch Calibration. We bold the best average result among different calibration techniques in both prompt and response classification.}
\label{app:adv_results}
\end{table}

%% file: appendix/app_adv_model_dep.tex
\begin{table}[htbp]
\centering
\begin{adjustbox}{max width=\textwidth}
\begin{tabular}{lccccccccccc}
\toprule[1pt]
\multirow{2}{*}{\textbf{Guard Model}} & \multicolumn{11}{c}{\textbf{Response Model}}  \\ 
                & \textbf{Baichuan2} & \textbf{Qwen} & \textbf{Solar} & \textbf{Llama2} & \textbf{Vicuna} & \textbf{Orca2} & \textbf{Koala} & \textbf{OpenChat} & \textbf{Starling} & \textbf{Zephyr} & \textbf{Avg.} \\ 
\midrule[0.5pt]
Llama-Guard & 26.9 & 23.0 & 49.4 & 10.5 & 28.0 & 26.4 & 27.4 & 40.3 & 46.3 & 38.5 & 31.7 \\
+ TS & 26.0 & 21.2 & 46.9 & 8.9 & 25.1 & 22.3 & 24.7 & 35.0 & 43.7 & 36.3 & 29.0 \\
+ CC & 25.5 & 17.1 & 46.3 & 10.0 & 26.3 & 22.2 & 31.8 & 35.1 & 40.1 & 31.5 & \textbf{28.6} \\
+ BC & 25.5 & 16.6 & 48.0 & 21.7 & 26.4 & 23.3 & 26.9 & 36.9 & 47.3 & 40.6 & 31.3 \\
\midrule[0.5pt]
Llama-Guard2 & 18.2 & 5.8 & 27.1 & 7.9 & 28.4 & 25.2 & 30.4 & 28.5 & 37.0 & 39.4 & 24.8 \\
+ TS & 15.4 & 5.3 & 25.0 & 7.1 & 25.9 & 22.9 & 27.5 & 26.0 & 33.7 & 36.5 & \textbf{22.5} \\
+ CC & 40.3 & 31.1 & 31.7 & 36.3 & 29.1 & 38.1 & 43.0 & 39.4 & 27.4 & 32.5 & 34.9 \\
+ BC & 18.3 & 7.8 & 29.7 & 15.6 & 28.3 & 25.2 & 28.6 & 28.5 & 38.6 & 40.5 & 26.1 \\
\midrule[0.5pt]
Llama-Guard3 & 33.7 & 17.1 & 31.0 & 27.4 & 20.5 & 27.3 & 36.5 & 34.2 & 27.6 & 23.1 & 27.8 \\
+ TS & 31.0 & 15.1 & 28.3 & 25.5 & 18.9 & 24.6 & 33.0 & 31.8 & 26.4 & 23.2 & \textbf{25.8} \\
+ CC & 46.1 & 30.8 & 42.6 & 45.7 & 36.9 & 39.8 & 48.0 & 43.4 & 27.9 & 29.0 & 39.0 \\
+ BC & 32.0 & 18.0 & 25.5 & 31.8 & 19.2 & 25.5 & 30.4 & 28.3 & 32.2 & 26.4 & 26.9 \\
\midrule[0.5pt]
Aegis-Guard-D & 35.5 & 27.1 & 22.2 & 40.8 & 34.0 & 33.9 & 31.3 & 30.9 & 27.8 & 30.9 & 31.4 \\
+ TS & 29.5 & 26.3 & 20.8 & 40.8 & 30.1 & 31.4 & 27.2 & 30.2 & 27.3 & 31.2 & 29.5 \\
+ CC & 27.7 & 17.3 & 26.5 & 25.9 & 28.8 & 25.7 & 23.1 & 29.4 & 34.3 & 34.2 & \textbf{27.3} \\
+ BC & 28.9 & 23.6 & 25.5 & 43.9 & 29.5 & 26.0 & 20.4 & 29.1 & 40.1 & 39.7 & 30.7 \\
\midrule[0.5pt]
Aegis-Guard-P & 22.8 & 17.4 & 28.3 & 23.6 & 26.2 & 28.2 & 32.1 & 35.5 & 36.3 & 25.7 & 27.6 \\
+ TS & 19.5 & 18.5 & 24.6 & 26.6 & 24.4 & 26.2 & 29.9 & 32.1 & 33.5 & 29.4 & \textbf{26.5} \\
+ CC & 23.7 & 18.2 & 27.5 & 25.4 & 26.6 & 28.9 & 32.9 & 35.1 & 35.3 & 24.6 & 27.8 \\
+ BC & 22.3 & 20.5 & 29.3 & 39.4 & 26.2 & 27.6 & 29.0 & 35.8 & 43.8 & 35.4 & 30.9 \\
\midrule[0.5pt]
HarmB-Llama & 17.7 & 6.4 & 25.0 & 23.5 & 16.0 & 19.5 & 26.7 & 23.2 & 23.3 & 20.1 & 20.1 \\
+ TS & 17.3 & 7.6 & 22.5 & 22.6 & 15.6 & 18.5 & 26.2 & 23.5 & 24.1 & 21.3 & \textbf{19.9} \\
+ CC & 32.0 & 23.9 & 31.8 & 24.8 & 23.7 & 28.8 & 37.8 & 29.2 & 21.9 & 21.5 & 27.5 \\
+ BC & 17.8 & 8.3 & 23.6 & 24.0 & 15.9 & 19.1 & 25.6 & 22.4 & 23.9 & 20.3 & 20.1 \\

\midrule[0.5pt]
HarmB-Mistral & 28.0 & 3.0 & 30.1 & 12.8 & 16.6 & 17.5 & 16.0 & 14.9 & 27.7 & 19.9 & 18.6 \\
+ TS & 26.4 & 5.6 & 27.3 & 12.7 & 15.2 & 14.3 & 15.6 & 14.1 & 26.1 & 19.5 & \textbf{17.7} \\
+ CC & 34.5 & 10.8 & 39.2 & 19.4 & 19.1 & 25.8 & 35.2 & 25.3 & 26.0 & 20.2 & 25.6 \\
+ BC & 27.1 & 2.9 & 23.8 & 17.5 & 16.6 & 17.1 & 11.4 & 14.0 & 27.0 & 19.6 & \textbf{17.7} \\
\midrule[0.5pt]
MD-Judge & 22.4 & 14.4 & 19.3 & 24.1 & 19.9 & 16.7 & 26.2 & 25.5 & 26.0 & 17.9 & \textbf{21.2} \\
+ TS & 22.2 & 17.5 & 19.6 & 27.4 & 20.3 & 17.2 & 23.2 & 21.7 & 29.4 & 20.1 & 21.9 \\
+ CC & 47.2 & 40.9 & 30.3 & 61.5 & 41.7 & 39.4 & 44.2 & 37.1 & 25.9 & 24.8 & 39.3 \\
+ BC & 20.7 & 17.2 & 21.5 & 38.5 & 20.1 & 15.7 & 17.6 & 26.8 & 39.0 & 30.3 & 24.7 \\
\midrule[0.5pt]
WildGuard & 22.1 & 9.2 & 15.5 & 17.0 & 11.2 & 20.1 & 37.3 & 25.4 & 18.6 & 21.3 & 19.8 \\
+ TS & 19.6 & 6.7 & 15.2 & 12.3 & 12.0 & 19.2 & 34.6 & 25.0 & 16.3 & 21.3 & \textbf{18.2} \\
+ CC & 24.3 & 12.2 & 20.1 & 21.3 & 16.1 & 20.9 & 39.2 & 31.3 & 20.6 & 24.0 & 23.0 \\
+ BC & 21.9 & 10.4 & 13.6 & 23.7 & 11.0 & 20.2 & 34.7 & 22.8 & 15.0 & 23.8 & 19.7 \\
\bottomrule[1pt]
\end{tabular}
\end{adjustbox}
\caption{ ECE (\%) $\downarrow$ performance comparison of different calibration techniques. For each guard model, we report the original calibration results in the first row and the rest results using TS: Temperature Scaling, CC: Contextual Calibration, BC: Batch Calibration, in the following three rows. We bold the best average result among different calibration techniques for each guard model.}
\label{app:adv_model_dep}
\end{table}

%% file: appendix/app_ts_valid.tex
\begin{table}[htbp]
\centering

\begin{adjustbox}{max width=\textwidth}
\begin{tabular}{lccc|ccc}
\toprule[1pt]
\multirow{2}{*}{\textbf{Model}} & \multicolumn{3}{c|}{\textbf{Prompt Classification}} & \multicolumn{3}{c}{\textbf{Response Classification}} \\ 
               & \textbf{Origin} & \textbf{TS(XSTest)} & \textbf{TS(In-Domain)} & \textbf{Origin} & \textbf{TS(XSTest)} & \textbf{TS(In-Domain)} \\ 
\midrule[0.5pt]
Llama-Guard    & 28.6 & 26.0 & 19.9 & 14.2 & 14.0 & 11.6   \\
Llama-Guard2   & 26.8 & 26.0 & 19.9 & 12.8 & 13.6 & 9.8  \\
Llama-Guard3   & 20.5 & 20.4 & 16.4 & 11.1 & 13.0 & 8.6  \\
\bottomrule[1pt]
\end{tabular}
\end{adjustbox}
\caption{ECE (\%) $\downarrow$ performance comparison of different validation sets.}
\label{app:ts_valid}
\end{table}

%% file: appendix/app_abl_length.tex
\begin{table}[htbp]
\centering

\begin{tabular}{l|ccc}
\toprule[1pt]
\textbf{Length (L) Range} & \textbf{0 $\leq$ L $<$ 200} & \textbf{200 $\leq$ L $<$ 500} & \textbf{500 $\leq$ L} \\
\midrule[0.5pt]
Llama-Guard  & 68.4 & 76.5 & 66.2 \\ 
Llama-Guard2 & 60.9 & 61.9 & 50.7 \\ 
Llama-Guard3 & 49.6 & 46.7 & 33.4 \\ 
\bottomrule[1pt]
\end{tabular}
\caption{ ECE (\%) $\downarrow$ performance comparison among different length ranges.}
\label{app:abl_length}
\end{table}

%% file: appendix/app_blackbox_response_model.tex
\begin{table}[htbp]
\centering
\begin{adjustbox}{max width=\textwidth}
\begin{tabular}{l|ccc}
\toprule[1pt]
\multirow{2}{*}{\textbf{Guard Model}} & \multicolumn{3}{c}{\textbf{Response Model}}  \\ 
               & \textbf{GPT-3.5} & \textbf{GPT-4} & \textbf{Claude-2}  \\ 
\midrule[0.5pt]
Llama-Guard      & 39.1 & 0.0  & 0.8  \\ 
Llama-Guard2     & 18.8 & 0.0  & 0.0  \\ 
Llama-Guard3     & 0.0  & 0.0  & 0.0  \\ 
Aegis-Guard-D    & 0.0  & 0.0  & 10.7 \\ 
Aegis-Guard-P    & 26.6 & 0.0  & 0.0  \\ 
HarmB-Llama      & 31.2 & 0.0  & 0.0  \\ 
HarmB-Mistral    & 30.5 & 0.0  & 0.0  \\ 
MD-Judge         & 21.5 & 0.0  & 3.2  \\ 
WildGuard        & 9.0  & 0.0  & 0.0  \\ 
\bottomrule[1pt]
\end{tabular}
\end{adjustbox}
\caption{ ECE (\%) $\downarrow$ performances of response classification on Harmbench-adv set on black-box response models.}
\label{app:model_comparison}
\end{table}

%% file: appendix/app_fpr_fnr.tex
\begin{table}[htbp]
\centering

\begin{adjustbox}{max width=\textwidth}
\begin{tabular}{lccccccc|cccccc}
\toprule[1pt]
\multirow{2}{*}{\textbf{Model}} & \multicolumn{7}{c|}{\textbf{Prompt Classification}} & \multicolumn{4}{c}{\textbf{Response Classification}} \\ 
               & \textbf{Metric} & \textbf{OAI} & \textbf{ToxiC} & \textbf{SimpST} & \textbf{Aegis} & \textbf{XST} & \textbf{HarmB} & \textbf{WildGT} & \textbf{BeaverT} & \textbf{S-RLHF} & \textbf{HarmB} & \textbf{WildGT} \\ 
\midrule[0.5pt]
\multirow{2}{*}{Llama-Guard}  & FPR & 8.4 & 1.4 & - & 3.2 & 15.2 & - & 2.4 & 8.5 & 16.6 & 12.9 & 0.8 \\
                              & FNR & 28.9 & 53.0 & 13.0 & 39.9 & 17.0 & 49.8 & 60.9 & 46.2 & 49.5 & 47.0 & 67.3 \\
\midrule[0.5pt]
\multirow{2}{*}{Llama-Guard2} & FPR & 8.2 & 3.1 & - & 4.8 & 7.6 & - & 3.6 & 7.9 & 21.3 & 19.3 & 3.1 \\
                              & FNR & 27.4 & 62.7 & 8.0 & 42.5 & 12.5 & 11.3 & 43.6 & 39.1 & 28.3 & 21.5 & 41.5 \\
\midrule[0.5pt]
\multirow{2}{*}{Llama-Guard3} & FPR & 9.2 & 5.0 & - & 2.4 & 2.8 & - & 4.4 & 4.0 & 17.7 & 35.3 & 3.5 \\
                              & FNR & 21.5 & 50.0 & 1.0 & 43.3 & 18.0 & 2.1 & 34.9 & 46.0 & 35.8 & 3.7 & 35.6 \\
\bottomrule[1pt]
\end{tabular}
\end{adjustbox}
\caption{FPR (\%) $\downarrow$ and FNR (\%) $\downarrow$ performances of prompt and response classification on existing public benchmarks. }
\label{app:fpr_fnr}
\end{table}

%% file: iclr2025_conference.bbl
\begin{thebibliography}{54}
\providecommand{\natexlab}[1]{#1}
\providecommand{\url}[1]{\texttt{#1}}
\expandafter\ifx\csname urlstyle\endcsname\relax
  \providecommand{\doi}[1]{doi: #1}\else
  \providecommand{\doi}{doi: \begingroup \urlstyle{rm}\Url}\fi

\bibitem[Abbas et~al.(2024)Abbas, Zhou, Ram, Baracaldo, Samulowitz, Salonidis, and Chen]{abbas2024enhancing}
Momin Abbas, Yi~Zhou, Parikshit Ram, Nathalie Baracaldo, Horst Samulowitz, Theodoros Salonidis, and Tianyi Chen.
\newblock Enhancing in-context learning via linear probe calibration.
\newblock In \emph{International Conference on Artificial Intelligence and Statistics}, pp.\  307--315. PMLR, 2024.

\bibitem[Anil et~al.(2023)Anil, Dai, Firat, Johnson, Lepikhin, Passos, Shakeri, Taropa, Bailey, Chen, et~al.]{palm2}
Rohan Anil, Andrew~M Dai, Orhan Firat, Melvin Johnson, Dmitry Lepikhin, Alexandre Passos, Siamak Shakeri, Emanuel Taropa, Paige Bailey, Zhifeng Chen, et~al.
\newblock Palm 2 technical report.
\newblock \emph{arXiv preprint arXiv:2305.10403}, 2023.

\bibitem[Brown(2020)]{brown2020language}
Tom~B Brown.
\newblock Language models are few-shot learners.
\newblock \emph{arXiv preprint arXiv:2005.14165}, 2020.

\bibitem[Chao et~al.(2024)Chao, Debenedetti, Robey, Andriushchenko, Croce, Sehwag, Dobriban, Flammarion, Pappas, Tramer, et~al.]{chao2024jailbreakbench}
Patrick Chao, Edoardo Debenedetti, Alexander Robey, Maksym Andriushchenko, Francesco Croce, Vikash Sehwag, Edgar Dobriban, Nicolas Flammarion, George~J Pappas, Florian Tramer, et~al.
\newblock Jailbreakbench: An open robustness benchmark for jailbreaking large language models.
\newblock \emph{arXiv preprint arXiv:2404.01318}, 2024.

\bibitem[Chen \& Yang(2023)Chen and Yang]{chen2023unlearn}
Jiaao Chen and Diyi Yang.
\newblock Unlearn what you want to forget: Efficient unlearning for llms.
\newblock \emph{arXiv preprint arXiv:2310.20150}, 2023.

\bibitem[Chen et~al.(2022)Chen, Yuan, Cui, Liu, and Ji]{chen2022close}
Yangyi Chen, Lifan Yuan, Ganqu Cui, Zhiyuan Liu, and Heng Ji.
\newblock A close look into the calibration of pre-trained language models.
\newblock \emph{arXiv preprint arXiv:2211.00151}, 2022.

\bibitem[Dai et~al.(2023)Dai, Pan, Sun, Ji, Xu, Liu, Wang, and Yang]{saferlhf}
Josef Dai, Xuehai Pan, Ruiyang Sun, Jiaming Ji, Xinbo Xu, Mickel Liu, Yizhou Wang, and Yaodong Yang.
\newblock Safe rlhf: Safe reinforcement learning from human feedback.
\newblock \emph{arXiv preprint arXiv:2310.12773}, 2023.

\bibitem[Dubey et~al.(2024)Dubey, Jauhri, Pandey, Kadian, Al-Dahle, Letman, Mathur, Schelten, Yang, Fan, et~al.]{llama3}
Abhimanyu Dubey, Abhinav Jauhri, Abhinav Pandey, Abhishek Kadian, Ahmad Al-Dahle, Aiesha Letman, Akhil Mathur, Alan Schelten, Amy Yang, Angela Fan, et~al.
\newblock The llama 3 herd of models.
\newblock \emph{arXiv preprint arXiv:2407.21783}, 2024.

\bibitem[Fei et~al.(2023)Fei, Hou, Chen, and Bosselut]{fei2023mitigating}
Yu~Fei, Yifan Hou, Zeming Chen, and Antoine Bosselut.
\newblock Mitigating label biases for in-context learning.
\newblock \emph{arXiv preprint arXiv:2305.19148}, 2023.

\bibitem[Ghosh et~al.(2024)Ghosh, Varshney, Galinkin, and Parisien]{aegis}
Shaona Ghosh, Prasoon Varshney, Erick Galinkin, and Christopher Parisien.
\newblock Aegis: Online adaptive ai content safety moderation with ensemble of llm experts.
\newblock \emph{arXiv preprint arXiv:2404.05993}, 2024.

\bibitem[Guo et~al.(2017)Guo, Pleiss, Sun, and Weinberger]{guo2017calibration}
Chuan Guo, Geoff Pleiss, Yu~Sun, and Kilian~Q Weinberger.
\newblock On calibration of modern neural networks.
\newblock In \emph{International conference on machine learning}, pp.\  1321--1330. PMLR, 2017.

\bibitem[Han et~al.(2024)Han, Rao, Ettinger, Jiang, Lin, Lambert, Choi, and Dziri]{wildguard}
Seungju Han, Kavel Rao, Allyson Ettinger, Liwei Jiang, Bill~Yuchen Lin, Nathan Lambert, Yejin Choi, and Nouha Dziri.
\newblock Wildguard: Open one-stop moderation tools for safety risks, jailbreaks, and refusals of llms.
\newblock \emph{arXiv preprint arXiv:2406.18495}, 2024.

\bibitem[Huang et~al.(2022)Huang, Gu, Wang, Xiao, Liu, and Wang]{ECBNN}
Hengguan Huang, Xiangming Gu, Hao Wang, Chang Xiao, Hongfu Liu, and Ye~Wang.
\newblock Extrapolative continuous-time bayesian neural network for fast training-free test-time adaptation.
\newblock In \emph{NeurIPS}, 2022.

\bibitem[Inan et~al.(2023)Inan, Upasani, Chi, Rungta, Iyer, Mao, Tontchev, Hu, Fuller, Testuggine, et~al.]{llamaguard}
Hakan Inan, Kartikeya Upasani, Jianfeng Chi, Rashi Rungta, Krithika Iyer, Yuning Mao, Michael Tontchev, Qing Hu, Brian Fuller, Davide Testuggine, et~al.
\newblock Llama guard: Llm-based input-output safeguard for human-ai conversations.
\newblock \emph{arXiv preprint arXiv:2312.06674}, 2023.

\bibitem[Ji et~al.(2024)Ji, Liu, Dai, Pan, Zhang, Bian, Chen, Sun, Wang, and Yang]{beavertails}
Jiaming Ji, Mickel Liu, Josef Dai, Xuehai Pan, Chi Zhang, Ce~Bian, Boyuan Chen, Ruiyang Sun, Yizhou Wang, and Yaodong Yang.
\newblock Beavertails: Towards improved safety alignment of llm via a human-preference dataset.
\newblock \emph{Advances in Neural Information Processing Systems}, 36, 2024.

\bibitem[Jiang et~al.(2021)Jiang, Araki, Ding, and Neubig]{jiang2021can}
Zhengbao Jiang, Jun Araki, Haibo Ding, and Graham Neubig.
\newblock How can we know when language models know? on the calibration of language models for question answering.
\newblock \emph{Transactions of the Association for Computational Linguistics}, 9:\penalty0 962--977, 2021.

\bibitem[Kadavath et~al.(2022)Kadavath, Conerly, Askell, Henighan, Drain, Perez, Schiefer, Hatfield-Dodds, DasSarma, Tran-Johnson, et~al.]{kadavath2022language}
Saurav Kadavath, Tom Conerly, Amanda Askell, Tom Henighan, Dawn Drain, Ethan Perez, Nicholas Schiefer, Zac Hatfield-Dodds, Nova DasSarma, Eli Tran-Johnson, et~al.
\newblock Language models (mostly) know what they know.
\newblock \emph{arXiv preprint arXiv:2207.05221}, 2022.

\bibitem[Lees et~al.(2022)Lees, Tran, Tay, Sorensen, Gupta, Metzler, and Vasserman]{perspective}
Alyssa Lees, Vinh~Q Tran, Yi~Tay, Jeffrey Sorensen, Jai Gupta, Donald Metzler, and Lucy Vasserman.
\newblock A new generation of perspective api: Efficient multilingual character-level transformers.
\newblock In \emph{Proceedings of the 28th ACM SIGKDD conference on knowledge discovery and data mining}, pp.\  3197--3207, 2022.

\bibitem[Li et~al.(2024)Li, Dong, Wang, Hu, Zuo, Lin, Qiao, and Shao]{mdjudge}
Lijun Li, Bowen Dong, Ruohui Wang, Xuhao Hu, Wangmeng Zuo, Dahua Lin, Yu~Qiao, and Jing Shao.
\newblock Salad-bench: A hierarchical and comprehensive safety benchmark for large language models.
\newblock \emph{arXiv preprint arXiv:2402.05044}, 2024.

\bibitem[Lin et~al.(2022)Lin, Hilton, and Evans]{lin2022teaching}
Stephanie Lin, Jacob Hilton, and Owain Evans.
\newblock Teaching models to express their uncertainty in words.
\newblock \emph{arXiv preprint arXiv:2205.14334}, 2022.

\bibitem[Lin et~al.(2023)Lin, Wang, Tong, Wang, Guo, Wang, and Shang]{toxicchat}
Zi~Lin, Zihan Wang, Yongqi Tong, Yangkun Wang, Yuxin Guo, Yujia Wang, and Jingbo Shang.
\newblock Toxicchat: Unveiling hidden challenges of toxicity detection in real-world user-ai conversation.
\newblock \emph{arXiv preprint arXiv:2310.17389}, 2023.

\bibitem[Liu \& Wang(2023)Liu and Wang]{liu2023towards}
Hongfu Liu and Ye~Wang.
\newblock Towards informative few-shot prompt with maximum information gain for in-context learning.
\newblock \emph{arXiv preprint arXiv:2310.08923}, 2023.

\bibitem[Liu et~al.(2024{\natexlab{a}})Liu, Xie, Wang, and Shieh]{liu2024advancing}
Hongfu Liu, Yuxi Xie, Ye~Wang, and Michael Shieh.
\newblock Advancing adversarial suffix transfer learning on aligned large language models.
\newblock \emph{arXiv preprint arXiv:2408.14866}, 2024{\natexlab{a}}.

\bibitem[Liu et~al.(2024{\natexlab{b}})Liu, Yao, Jia, Casper, Baracaldo, Hase, Xu, Yao, Li, Varshney, et~al.]{liu2024rethinking}
Sijia Liu, Yuanshun Yao, Jinghan Jia, Stephen Casper, Nathalie Baracaldo, Peter Hase, Xiaojun Xu, Yuguang Yao, Hang Li, Kush~R Varshney, et~al.
\newblock Rethinking machine unlearning for large language models.
\newblock \emph{arXiv preprint arXiv:2402.08787}, 2024{\natexlab{b}}.

\bibitem[Liu et~al.(2023)Liu, Xu, Chen, and Xiao]{autodan}
Xiaogeng Liu, Nan Xu, Muhao Chen, and Chaowei Xiao.
\newblock Autodan: Generating stealthy jailbreak prompts on aligned large language models.
\newblock \emph{arXiv preprint arXiv:2310.04451}, 2023.

\bibitem[Markov et~al.(2023)Markov, Zhang, Agarwal, Nekoul, Lee, Adler, Jiang, and Weng]{openai}
Todor Markov, Chong Zhang, Sandhini Agarwal, Florentine~Eloundou Nekoul, Theodore Lee, Steven Adler, Angela Jiang, and Lilian Weng.
\newblock A holistic approach to undesired content detection in the real world.
\newblock In \emph{Proceedings of the AAAI Conference on Artificial Intelligence}, volume~37, pp.\  15009--15018, 2023.

\bibitem[Mazeika et~al.(2024)Mazeika, Phan, Yin, Zou, Wang, Mu, Sakhaee, Li, Basart, Li, et~al.]{harmbench}
Mantas Mazeika, Long Phan, Xuwang Yin, Andy Zou, Zifan Wang, Norman Mu, Elham Sakhaee, Nathaniel Li, Steven Basart, Bo~Li, et~al.
\newblock Harmbench: A standardized evaluation framework for automated red teaming and robust refusal.
\newblock \emph{arXiv preprint arXiv:2402.04249}, 2024.

\bibitem[Mielke et~al.(2022)Mielke, Szlam, Dinan, and Boureau]{mielke2022reducing}
Sabrina~J Mielke, Arthur Szlam, Emily Dinan, and Y-Lan Boureau.
\newblock Reducing conversational agents’ overconfidence through linguistic calibration.
\newblock \emph{Transactions of the Association for Computational Linguistics}, 10:\penalty0 857--872, 2022.

\bibitem[Minderer et~al.(2021)Minderer, Djolonga, Romijnders, Hubis, Zhai, Houlsby, Tran, and Lucic]{minderer2021revisiting}
Matthias Minderer, Josip Djolonga, Rob Romijnders, Frances Hubis, Xiaohua Zhai, Neil Houlsby, Dustin Tran, and Mario Lucic.
\newblock Revisiting the calibration of modern neural networks.
\newblock \emph{Advances in Neural Information Processing Systems}, 34:\penalty0 15682--15694, 2021.

\bibitem[Naeini et~al.(2015)Naeini, Cooper, and Hauskrecht]{naeini2015obtaining}
Mahdi~Pakdaman Naeini, Gregory Cooper, and Milos Hauskrecht.
\newblock Obtaining well calibrated probabilities using bayesian binning.
\newblock In \emph{Proceedings of the AAAI conference on artificial intelligence}, volume~29, 2015.

\bibitem[Nguyen \& O'Connor(2015)Nguyen and O'Connor]{posteriorllm}
Khanh Nguyen and Brendan O'Connor.
\newblock Posterior calibration and exploratory analysis for natural language processing models.
\newblock \emph{arXiv preprint arXiv:1508.05154}, 2015.

\bibitem[Ouyang et~al.(2022)Ouyang, Wu, Jiang, Almeida, Wainwright, Mishkin, Zhang, Agarwal, Slama, Ray, et~al.]{rlhf}
Long Ouyang, Jeffrey Wu, Xu~Jiang, Diogo Almeida, Carroll Wainwright, Pamela Mishkin, Chong Zhang, Sandhini Agarwal, Katarina Slama, Alex Ray, et~al.
\newblock Training language models to follow instructions with human feedback.
\newblock \emph{Advances in neural information processing systems}, 35:\penalty0 27730--27744, 2022.

\bibitem[Rafailov et~al.(2024)Rafailov, Sharma, Mitchell, Manning, Ermon, and Finn]{dpo}
Rafael Rafailov, Archit Sharma, Eric Mitchell, Christopher~D Manning, Stefano Ermon, and Chelsea Finn.
\newblock Direct preference optimization: Your language model is secretly a reward model.
\newblock \emph{Advances in Neural Information Processing Systems}, 36, 2024.

\bibitem[Rebedea et~al.(2023)Rebedea, Dinu, Sreedhar, Parisien, and Cohen]{nemo}
Traian Rebedea, Razvan Dinu, Makesh Sreedhar, Christopher Parisien, and Jonathan Cohen.
\newblock Nemo guardrails: A toolkit for controllable and safe llm applications with programmable rails.
\newblock \emph{arXiv preprint arXiv:2310.10501}, 2023.

\bibitem[R{\"o}ttger et~al.(2023)R{\"o}ttger, Kirk, Vidgen, Attanasio, Bianchi, and Hovy]{xstest}
Paul R{\"o}ttger, Hannah~Rose Kirk, Bertie Vidgen, Giuseppe Attanasio, Federico Bianchi, and Dirk Hovy.
\newblock Xstest: A test suite for identifying exaggerated safety behaviours in large language models.
\newblock \emph{arXiv preprint arXiv:2308.01263}, 2023.

\bibitem[Schmidt \& Wiegand(2017)Schmidt and Wiegand]{hatespeech}
Anna Schmidt and Michael Wiegand.
\newblock A survey on hate speech detection using natural language processing.
\newblock In \emph{Proceedings of the fifth international workshop on natural language processing for social media}, pp.\  1--10, 2017.

\bibitem[Touvron et~al.(2023)Touvron, Martin, Stone, Albert, Almahairi, Babaei, Bashlykov, Batra, Bhargava, Bhosale, et~al.]{llama2}
Hugo Touvron, Louis Martin, Kevin Stone, Peter Albert, Amjad Almahairi, Yasmine Babaei, Nikolay Bashlykov, Soumya Batra, Prajjwal Bhargava, Shruti Bhosale, et~al.
\newblock Llama 2: Open foundation and fine-tuned chat models.
\newblock \emph{arXiv preprint arXiv:2307.09288}, 2023.

\bibitem[Vidgen et~al.(2023)Vidgen, Kirk, Qian, Scherrer, Kannappan, Hale, and R{\"o}ttger]{simplesafetytests}
Bertie Vidgen, Hannah~Rose Kirk, Rebecca Qian, Nino Scherrer, Anand Kannappan, Scott~A Hale, and Paul R{\"o}ttger.
\newblock Simplesafetytests: a test suite for identifying critical safety risks in large language models.
\newblock \emph{arXiv preprint arXiv:2311.08370}, 2023.

\bibitem[Wang et~al.(2023)Wang, Chen, Pei, Xie, Kang, Zhang, Xu, Xiong, Dutta, Schaeffer, et~al.]{wang2023decodingtrust}
Boxin Wang, Weixin Chen, Hengzhi Pei, Chulin Xie, Mintong Kang, Chenhui Zhang, Chejian Xu, Zidi Xiong, Ritik Dutta, Rylan Schaeffer, et~al.
\newblock Decodingtrust: A comprehensive assessment of trustworthiness in gpt models.
\newblock In \emph{NeurIPS}, 2023.

\bibitem[Wang \& Yeung(2016)Wang and Yeung]{BDL}
Hao Wang and Dit-Yan Yeung.
\newblock Towards bayesian deep learning: A framework and some existing methods.
\newblock \emph{TDKE}, 28\penalty0 (12):\penalty0 3395--3408, 2016.

\bibitem[Wang \& Yeung(2020)Wang and Yeung]{BDLSurvey}
Hao Wang and Dit-Yan Yeung.
\newblock A survey on bayesian deep learning.
\newblock \emph{CSUR}, 53\penalty0 (5):\penalty0 1--37, 2020.

\bibitem[Wang et~al.(2016)Wang, Xingjian, and Yeung]{NPN}
Hao Wang, SHI Xingjian, and Dit-Yan Yeung.
\newblock Natural-parameter networks: A class of probabilistic neural networks.
\newblock In \emph{NIPS}, pp.\  118--126, 2016.

\bibitem[Wang et~al.(2020)Wang, He, and Katabi]{CIDA}
Hao Wang, Hao He, and Dina Katabi.
\newblock Continuously indexed domain adaptation.
\newblock In \emph{ICML}, 2020.

\bibitem[Wang et~al.(2024)Wang, Shi, Han, Metaxas, and Wang]{BLoB}
Yibin Wang, Haizhou Shi, Ligong Han, Dimitris Metaxas, and Hao Wang.
\newblock Blob: Bayesian low-rank adaptation by backpropagation for large language models.
\newblock In \emph{NeurIPS}, 2024.

\bibitem[Xiong et~al.(2023)Xiong, Hu, Lu, Li, Fu, He, and Hooi]{xiong2023can}
Miao Xiong, Zhiyuan Hu, Xinyang Lu, Yifei Li, Jie Fu, Junxian He, and Bryan Hooi.
\newblock Can llms express their uncertainty? an empirical evaluation of confidence elicitation in llms.
\newblock \emph{arXiv preprint arXiv:2306.13063}, 2023.

\bibitem[Xu et~al.(2024)Xu, Jiang, Niu, Jia, Lin, and Poovendran]{xu2024safedecoding}
Zhangchen Xu, Fengqing Jiang, Luyao Niu, Jinyuan Jia, Bill~Yuchen Lin, and Radha Poovendran.
\newblock Safedecoding: Defending against jailbreak attacks via safety-aware decoding.
\newblock \emph{arXiv preprint arXiv:2402.08983}, 2024.

\bibitem[Xu et~al.(2022)Xu, Lee, Wang, Wang, et~al.]{GRDA}
Zihao Xu, Guang-He Lee, Yuyang Wang, Hao Wang, et~al.
\newblock Graph-relational domain adaptation.
\newblock In \emph{ICLR}, 2022.

\bibitem[Yuan et~al.(2024)Yuan, Xiong, Zeng, Yu, Jia, Song, and Li]{rigor}
Zhuowen Yuan, Zidi Xiong, Yi~Zeng, Ning Yu, Ruoxi Jia, Dawn Song, and Bo~Li.
\newblock Rigorllm: Resilient guardrails for large language models against undesired content.
\newblock \emph{arXiv preprint arXiv:2403.13031}, 2024.

\bibitem[Zeng et~al.(2024)Zeng, Liu, Mullins, Peran, Fernandez, Harkous, Narasimhan, Proud, Kumar, Radharapu, et~al.]{shieldgemma}
Wenjun Zeng, Yuchi Liu, Ryan Mullins, Ludovic Peran, Joe Fernandez, Hamza Harkous, Karthik Narasimhan, Drew Proud, Piyush Kumar, Bhaktipriya Radharapu, et~al.
\newblock Shieldgemma: Generative ai content moderation based on gemma.
\newblock \emph{arXiv preprint arXiv:2407.21772}, 2024.

\bibitem[Zhao et~al.(2021)Zhao, Wallace, Feng, Klein, and Singh]{zhao2021calibrate}
Zihao Zhao, Eric Wallace, Shi Feng, Dan Klein, and Sameer Singh.
\newblock Calibrate before use: Improving few-shot performance of language models.
\newblock In \emph{International conference on machine learning}, pp.\  12697--12706. PMLR, 2021.

\bibitem[Zhou et~al.(2023{\natexlab{a}})Zhou, Wan, Proleev, Mincu, Chen, Heller, and Roy]{zhou2023batch}
Han Zhou, Xingchen Wan, Lev Proleev, Diana Mincu, Jilin Chen, Katherine Heller, and Subhrajit Roy.
\newblock Batch calibration: Rethinking calibration for in-context learning and prompt engineering.
\newblock \emph{arXiv preprint arXiv:2309.17249}, 2023{\natexlab{a}}.

\bibitem[Zhou et~al.(2023{\natexlab{b}})Zhou, Wan, Vuli{\'c}, and Korhonen]{zhou2023survival}
Han Zhou, Xingchen Wan, Ivan Vuli{\'c}, and Anna Korhonen.
\newblock Survival of the most influential prompts: Efficient black-box prompt search via clustering and pruning.
\newblock \emph{arXiv preprint arXiv:2310.12774}, 2023{\natexlab{b}}.

\bibitem[Zhu et~al.(2023)Zhu, Xu, Wang, Zhang, and Mao]{calibrationllm}
Chiwei Zhu, Benfeng Xu, Quan Wang, Yongdong Zhang, and Zhendong Mao.
\newblock On the calibration of large language models and alignment.
\newblock \emph{arXiv preprint arXiv:2311.13240}, 2023.

\bibitem[Zou et~al.(2023)Zou, Wang, Carlini, Nasr, Kolter, and Fredrikson]{gcg}
Andy Zou, Zifan Wang, Nicholas Carlini, Milad Nasr, J~Zico Kolter, and Matt Fredrikson.
\newblock Universal and transferable adversarial attacks on aligned language models.
\newblock \emph{arXiv preprint arXiv:2307.15043}, 2023.

\end{thebibliography}
